%% file: DORL-final.tex
\documentclass[sigconf,natbib=true]{acmart}
\input{sections/definition}
\acmSubmissionID{4828}

\begin{document}

% \title{Alleviating Matthew Effect of Model-based Offline Reinforcement Learning in Recommendation}
% \title{Debiased Offline Reinforement Learning for Recommendation}
\title{Alleviating Matthew Effect of Offline Reinforcement Learning\\ in Interactive Recommendation}
% \title{KuaiRand: A Sequential Recommendation Dataset with Millions of Random Exposure Records}

\input{sections/author}

\begin{abstract}
% Pursuing users' long-term satisfaction is one of the main goals of recommender systems. Reinforcement learning (RL) is well suited for this task since RL considers how an intelligent agent should automatically make decisions within a specific context to pursue a long-term goal. 
% However, the biggest obstacle to RL's widespread adoption in recommendation is its online learning paradigm, i.e., interacting online with users is impractical. 
% Therefore, researchers resort to offline RL, whose objective is to let the learned policy perform better than the behavior policies, i.e., old policies in offline data. 
% The core challenge is the value overestimation problem, where values of out-of-distribution state-action pairs are optimistically estimated. To solve this, almost all offline RL algorithms consider conservatism, e.g., by constraining the learned policy to be close to behavior policies or by punishing the rarely visited state-action pairs. However, when employing offline RL in recommendation, this philosophy will cause a severe Matthew effect: popular items or categories get more opportunities to be recommended. % It hurts user experience.

Offline reinforcement learning (RL), a technology that offline learns a policy from logged data without the need to interact with online environments, has become a favorable choice in decision-making processes like interactive recommendation. 
%has drawn growing attention. 
Offline RL faces the value overestimation problem.
% , which is induced by the distributional shift in offline data. 
To address it, existing methods employ conservatism, e.g., by constraining the learned policy to be close to behavior policies or punishing the rarely visited state-action pairs. 
However, when applying such offline RL to recommendation, it will cause a severe Matthew effect, i.e., 
the rich get richer and the poor get poorer, by promoting popular items or categories while suppressing the less popular ones.
%popular items or categories get more opportunities to be recommended 
It is a notorious issue that needs to be addressed in practical recommender systems. 
%We conduct field studies to show how it hurts user experience.

In this paper, we aim to alleviate the Matthew effect in offline RL-based recommendation. Through theoretical analyses, we find that the conservatism of existing methods fails in pursuing users' long-term satisfaction. 
% It inspires us to add a penalty term to make the policy emphasize the data induced by the behavior policies with high entropy, 
% which re-introduces the exploration mechanism that the conservatism of offline RL has suppressed. 
It inspires us to add a penalty term to relax the pessimism on states with high entropy of the logging policy and indirectly penalizes actions leading to less diverse states.
This leads to the main technical contribution of the work: \textit{Debiased model-based Offline RL} (DORL) method. Experiments show that DORL not only captures user interests well but also alleviates the Matthew effect. 
The implementation is available via \textcolor{magenta}{\url{https://github.com/chongminggao/DORL-codes}}.
%can well capture user interest, meanwhile alleviating the Matthew effect. 
% We implement the model in an interactive recommendation setting where we can evaluate users' long-term satisfaction. 
% The experiments show the effectiveness of the proposed solution.

% Since offline RL faces value overestimation problem, the main 

\end{abstract}

\input{sections/meta}


\maketitle
\input{sections/01.introduction}
\input{sections/02.related}

\input{sections/03.preliminaries}

\input{sections/04.method}

\input{sections/05.experiment}

\section{Conclusion}
We point out that conservatism in offline RL can incur the Matthew effect in recommendation. We conduct  studies to show that the Matthew effect hurts users' long-term experiences in both the music and video datasets. Through theoretical analysis of the model-based RL framework, we show that the reason for amplifying the Matthew effect is the philosophy of suppressing uncertain samples. It inspires us to add a penalty term to make the policy emphasize the data induced by the behavior policies with high entropy. This will reintroduce the exploration mechanism that conservatism has suppressed, which alleviates the Matthew effect.

In the future, when fitting user interests is not a bottleneck anymore, researchers could consider higher-level goals, such as pursuing users' long-term satisfaction \cite{10.1145/3534678.3539040} or optimizing social utility \cite{Pareto22wsdm}. 
With the increase in high-quality offline data, we believe that offline RL can be better adapted to recommender systems to achieve these goals. During this process, many interesting yet challenging issues (such as the Matthew effect in this work) will be raised. After addressing these issues, we can create more intelligent recommender systems that benefit society.

\section*{Acknowledgements}
This work is supported by the National Key Research and Development Program of China (2021YFF0901603), the National Natural Science Foundation of China (61972372, U19A2079, 62121002), and the CCCD Key Lab of Ministry of Culture and Tourism.

\bibliographystyle{ACM-Reference-Format}
\bibliography{rethink}

\end{document}

%% file: sections/definition.tex
\AtBeginDocument{%
  }

\setcopyright{acmcopyright}
\copyrightyear{2018}
\acmYear{2018}
\acmDOI{XXXXXXX.XXXXXXX}

\acmConference[Conference acronym 'XX]{Make sure to enter the correct
  conference title from your rights confirmation emai}{June 03--05,
  2018}{Woodstock, NY}
\acmPrice{15.00}
\acmISBN{978-1-4503-XXXX-X/18/06}

\usepackage{booktabs}
\usepackage{multirow}
\usepackage{url}
\usepackage{enumitem}   
\usepackage{float} % For figure floating in appendix

\definecolor{myred}{rgb}{0.68627451, 0.14117647, 0.09803922}

\newcommand{\myeq}[1]{\hyperref[eq:#1]{Eq.~(\ref*{eq:#1})}}
\newcommand{\mysec}[1]{\hyperref[sec:#1]{Section~\ref*{sec:#1}}}
\newcommand{\mytable}[1]{\hyperref[tab:#1]{Table~\ref*{tab:#1}}}
\newcommand{\myfig}[1]{\hyperref[fig:#1]{Fig.~\ref*{fig:#1}}}
\newcommand{\myappendix}[1]{\hyperref[appendix:#1]{Appendix~\ref*{appendix:#1}}}
\newcommand{\myalg}[1]{\hyperref[alg:#1]{Algorithm~\ref*{alg:#1}}}
\newcommand{\myhyp}[1]{\hyperref[hyp:#1]{\textsc{Hypothesis}~\ref*{hyp:#1}}}

\newcommand{\mytheorem}[1]{\hyperref[theorem:#1]{Theorem~\ref*{theorem:#1}}}
\newcommand{\mylemma}[1]{\hyperref[lemma:#1]{Lemma~\ref*{lemma:#1}}}

%% file: sections/author.tex
\author{Chongming Gao}
\email{chongming.gao@gmail.com}
\orcid{0000-0002-5187-9196}
\affiliation{%
  \institution{University of Science and Technology of China}
  % \city{Hefei}
  % \state{Anhui}
  \country{}
}

\author{Kexin Huang}
\email{huangkx@mail.ustc.edu.cn}
\orcid{0009-0001-4868-0952}
\affiliation{%
  \institution{University of Science and Technology of China}
  % \city{Hefei}
  % \state{Anhui}
  \country{}
}

\author{Jiawei Chen}
\authornote{Corresponding author.}
\orcid{0000-0002-4752-2629}
\email{sleepyhunt@zju.edu.cn}
\affiliation{%
  \institution{Zhejiang University}
  % \streetaddress{1 Th{\o}rv{\"a}ld Circle}
  \city{Hangzhou}
  \country{China}
}

\author{Yuan Zhang}
\email{yuanz.pku@gmail.com}
\orcid{0000-0002-7849-208X}
\affiliation{%
  \institution{Kuaishou Technology Co., Ltd.}
  \country{}
}

\author{Biao Li}
\email{libiao@kuaishou.com}
\orcid{0000-0001-5667-5347}
\affiliation{%
  \institution{Kuaishou Technology Co., Ltd.}
  \country{}
}

\author{Peng Jiang}
\email{jiangpeng@kuaishou.com}
\orcid{0000-0002-9266-0780}
\affiliation{%
  \institution{Kuaishou Technology Co., Ltd.}
  \country{}
}

\author{Shiqi Wang}
\email{shiqi@cqu.edu.cn}
\orcid{0000-0002-5369-884X}
\affiliation{%
  \institution{Chongqing University}
  \city{Chongqing}
  \country{China}
}

\author{Zhong Zhang}
\email{zhongzhang@std.uestc.edu.cn}
\orcid{0000-0003-1349-9755}
\affiliation{%
  \institution{University of Electronic Science and Technology of China}
  \country{}
}

\author{Xiangnan He}
\authornotemark[1]
\email{xiangnanhe@gmail.com}
\orcid{0000-0001-8472-7992}
\affiliation{%
  \institution{University of Science and Technology of China}
  \country{}
}

\renewcommand{\shortauthors}{Chongming Gao et al.}

%% file: sections/meta.tex
% \copyrightyear{2022}
% \acmYear{2022}
% \setcopyright{acmcopyright}
% \acmConference[SIGIR '23] {Proceedings of the 46th International ACM SIGIR Conference on Research and Development in Information Retrieval}{July 23--27, 2023}{Taipei.}
% \acmBooktitle{Proceedings of the 46th International ACM SIGIR Conference on Research and Development in Information Retrieval (SIGIR '23), July 23--27, 2023, Taipei.}
% \acmPrice{15.00}
% \acmISBN{978-1-4503-9236-5/22/10}
% \acmDOI{10.1145/3511808.3557624}

% SIGIR 23
\copyrightyear{2023} 
\acmYear{2023} 
\setcopyright{acmlicensed}\acmConference[SIGIR '23]{Proceedings of the 46th
International ACM SIGIR Conference on Research and Development in
Information Retrieval}{July 23--27, 2023}{Taipei, Taiwan}
\acmBooktitle{Proceedings of the 46th International ACM SIGIR Conference on
Research and Development in Information Retrieval (SIGIR '23), July 23--27,
2023, Taipei, Taiwan}
\acmPrice{15.00}
\acmDOI{10.1145/3539618.3591636}
\acmISBN{978-1-4503-9408-6/23/07}

\begin{CCSXML}
<ccs2012>
<concept>
<concept_id>10002951.10003317.10003347.10003350</concept_id>
<concept_desc>Information systems~Recommender systems</concept_desc>
<concept_significance>500</concept_significance>
</concept>
</ccs2012>
\end{CCSXML}
\ccsdesc[500]{Information systems~Recommender systems}

%% file: sections/01.introduction.tex
\section{Introduction}

Recommender systems, a powerful tool for helping users select preferred items from massive items, are continuously investigated by e-commerce companies. 
Previously, researchers tried to dig up static user interests from historical data by developing supervised learning-based recommender models. With the recent development of deep learning and the rapid growth of available data, fitting user interests is not a bottleneck for now.

A desired recommendation policy should be able to satisfy users for a long time \cite{surrogate}. Therefore, it is natural to involve Reinforcement Learning (RL) which is a type of Machine Learning concerned with how an intelligent agent can take actions to pursue a long-term goal \cite{afsar2022reinforcement,cai2023reinforcing,liu2023exploration,cai2023two}. In this setting, the recommendation process is formulated as a sequential decision process where the recommender interacts with users and receives users' online feedback (i.e., rewards) to optimize users' long-term engagement, rather than fitting a model on a set of samples based on supervised learning \cite{minmin22,lixin19,ResAct}.
However, it is expensive and impractical to learn a policy from scratch with real users, which becomes the main obstacle that impedes the deployment of RL to recommender systems.

One remedy is to leverage historical interaction sequences, i.e., recommendation logs, to conduct offline RL (also called batch RL) \cite{offlineRLsurvey}. The objective is to learn an online policy that makes counterfactual decisions to perform better than the behavior policies induced by the offline data. However, without real-time feedback, directly employing conventional online RL algorithms in offline scenarios will result in poor performance due to the value overestimation problem in offline RL. The problem is induced when the function approximator of the agent tries to extrapolate values (e.g., Q-values in Q-learning \cite{watkins1992q}) for the state-action pairs that are not well-covered by logged data.
More specifically, since the RL model usually maximizes the expected value or trajectory reward, it will intrinsically prefer overestimated values induced by the extrapolation error, and the error will be compounded in the bootstrapping process when estimating Q-values, which results in unstable learning and divergence \cite{BEAR}. 
In recommendation, this may lead to an overestimation of user preferences for items that infrequently appear in the offline logs.

This is the core challenge for offline RL algorithms because of the inevitable mismatch between the offline dataset and the learned policy \cite{wang2020statistical}. 
To solve this problem, offline RL algorithms incorporate conservatism into the policy design. Model-free offline RL algorithms directly incorporate conservatism by constraining the learned policy to be close to the behavior policy \cite{BCQ,BEAR}, or by penalizing the learned value functions from being over-optimistic upon out-of-distribution (OOD) decisions \cite{CQL}. Model-based offline RL algorithms learn a pessimistic model as a proxy of the environment, which results in a conservative policy \cite{MOReL,MOPO}. This philosophy guarantees that offline RL models can stick to offline data without making OOD actions, which has been proven to be effective in lots of domains, such as robotic control \cite{ebert2018visual} or games \cite{agarwal2020optimistic}.

However, applying conservatism to recommender systems gives rise to a severe Matthew effect \cite{ExamingMatthew}, which can be summarized as ``the rich get richer and the poor get poorer''. In recommendation, it means that the popular items or categories in previous data will get larger opportunities to be recommended later, whereas the unpopular ones get neglected. This is catastrophic since users desire diverse recommendations and the repetition of certain contents will incur the filter bubble issue, which in turn hurts users' satisfaction even though users favored them before \cite{gao2022cirs,FBsigir22,recsys21best,zhang2023divide}. We will show the Matthew effect in the existing offline RL-based recommender (\myfig{conservative}), and analyze how users' satisfaction will be hurt (\myfig{effect}).

In this paper, we embrace the model-based RL paradigm. The basic idea is to learn a user model (i.e., world model) that captures users' preferences, then use it as a pseudo-environment (i.e., simulated users) to produce rewards to train a recommendation policy. 
Compared to model-free RL, model-based RL has several advantages in recommendation. First and foremost, model-based RL is much more sample efficient \cite{MOPO}. That it needs significantly fewer samples makes it more suitable for the highly sparse recommendation data. Second, explicitly learning the user model simplifies the problem and makes it easier to incorporate expert knowledge. For example, the user model can be implemented as any state-of-the-art recommendation model (e.g., DeepFM \cite{DeepFM} in this work) or sophisticated generative adversarial frameworks \cite{ResAct,xiangyuwww21}. 
Although some works have adopted this paradigm in their recommender systems \cite{gao2022cirs,jinhuang22,Keeping-recsys,Pseudo-Dyna-Q}, they did not explicitly consider the value overestimation problem in offline RL, not to mention the Matthew effect in the solutions.

To address the value overestimation problem while reducing the Matthew effect, we propose a \emph{Debiased model-based Offline RL} (DORL) method for recommendation. By theoretically analyzing the mismatch between real users' long-term satisfaction and the preferences estimated from the offline data, DORL adds a penalty term that relaxes the pessimism on states with high entropy of the logging policy and indirectly penalizes actions leading to less diverse states. 
% It makes the policy emphasize the data induced by the behavior policies with high entropy, e.g, a uniform distribution. 
By introducing such a counterfactual exploration mechanism, DORL can alleviate the Matthew effect in final recommendations.
Our contributions are summarized as:
\begin{itemize}
	\item We point out that conservatism in offline RL can incur the Matthew effect in recommendation. We show this phenomenon in existing methods and how it hurts user satisfaction.
	\item After theoretically analyzing how existing methods fail in recommendation, we propose the DORL model that introduces a counterfactual exploration in offline data.
	\item We demonstrate the effectiveness of DORL in an interactive recommendation setting, where alleviating the Matthew effect increases users' long-term experience.
\end{itemize}
% The interactive recommendation setting where we can evaluate users' long-term satisfaction. The results show the effectiveness of the proposed solution.

% By penalizing not only the uncertainty but also the entropy of the behavior policy, the learned policy can well capture user interest meanwhile alleviating the Matthew effect in the results. 

%% file: sections/02.related.tex
\vspace{-2mm}
\section{Related Work}
\label{sec:related}

Here, we briefly review the Matthew effect in recommendation. We introduce the interactive recommendation and offline RL.

\subsection{Matthew Effect in Recommendation}
% The Matthew effect in recommendation refers to the phenomenon that recommenders incline to recommend popular items, i.e. ``the rich get richer and the poor get poorer''. 
\citet{ExamingMatthew} confirmed the existence of the Matthew effect in YouTube's recommendation system, and \citet{QuantifyMatthew} gave a quantitative analysis of the Matthew effect in collaborative filtering-based recommenders.
A common way of mitigating the Matthew effect in recommendation is to take into account diversity \cite{www20Algorithmic,DGCN,diversitySigir21,wsdm21diversity}.
Another perspective on this problem is to remove popularity bias \cite{DICE}.
We consider the Matthew effect in offline RL-based recommendation systems. we will analyze why this problem occurs and provide a novel way to address it.

\vspace{-2mm}
\subsection{Interactive Recommendation}
\label{sec:IRS}

The interactive recommendation is a setting where a model interacts with a user online \cite{10.1145/3477495.3531969,gao2021advances,gao2021tutorial}. The model recommends items to the user and receives the user's real-time feedback. This process is repeated until the user quits. The model will update its policy with the goal to maximize the cumulative satisfaction over the whole interaction process (instead of learning on I.I.D. samples).
This setting well reflects the real-world recommendation scenarios, for example, a user will continuously watch short videos and leave feedback (e.g. click, add to favorite) until he chooses to quit. 

\begin{figure}[t!]
    \includegraphics[width=1\linewidth]{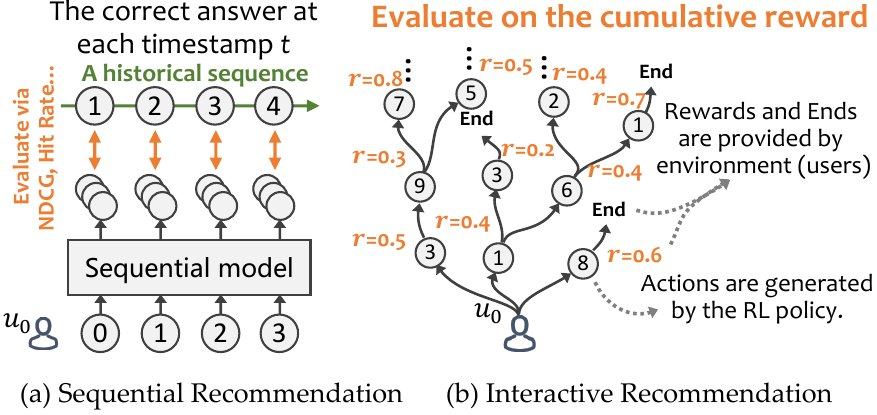}
    \vspace{-2mm}
    \caption{Illustration of evaluations in the traditional sequential recommendation and interactive recommendation settings.}
    \vspace{-2mm}
    \label{fig:settings}
\end{figure}
Here, we emphasize the most notable difference between the interactive recommendation setting and traditional sequential recommendation settings \cite{xinxin22,xinxin20}. \myfig{settings} illustrates the learning and evaluation processes in sequential and interactive recommendation settings. Sequential recommendation uses the philosophy of supervised learning, i.e., evaluating the top-$k$ results by comparing them with a set of ``correct'' answers in the test set and computing metrics such as Precision, Recall, NDCG, and Hit Rate. By contrast, interactive recommendation evaluates the results by accumulating the rewards along the interaction trajectories. There is no standard answer in interactive recommendation, which is challenging \cite{deffayet2023offline}. 

Interactive recommendation requires offline data of high quality, which hampers the development of this field for a long time.
We overcome this problem by using the recently-proposed datasets that support interactive learning and off-policy evaluation \cite{gilotte2018offline}.

\subsection{Offline Reinforcement Learning}
% Reinforcement Learning (RL) is a type of Machine Learning that has shown effectiveness in a lot of domains, such as Game AI \cite{silver2016mastering}, robotic manipulation \cite{gu2017deep}, autonomous driving \cite{kiran2021deep}, healthcare \cite{yu2021reinforcement}, and recommendation \cite{afsar2022reinforcement}.
% The process of this problem is typically defined as a Markov Decision Process (MDP): $\mathcal{M}=(\mathcal{S,A,T},d_0,r,\gamma)$, including state space $\mathcal{S}$, action space $\mathcal{A}$, transition function $\mathcal{T}$, initial state distribution $d_0$, reward function $r(s,a)$, and discount factor $\gamma$. 
% Though effective, the biggest obstacle to RL-based algorithms' widespread adoption is that online interaction can be impractical in real-world scenarios. It is imperative to introduce offline RL \cite{offlineRLsurvey}. The objective of offline RL is to learn from the offline data induced by old policies and perform better. The biggest challenge in offline RL is the value overestimation problem where values will be optimistically estimated on out-of-distribution (OOD) areas \cite{BEAR}.

Recently, many offline RL models have been proposed to overcome the value overestimation problem. 
For model-free methods, BCQ \cite{BCQ} uses a generative model to constrain probabilities of state-action pairs the policy utilizes, thus avoiding using rarely visited data to update the value network; CQL \cite{CQL} contains a conservative strategy to penalize the overestimated Q-values for the state-action pairs that have not appeared in the offline data; GAIL \cite{GAIL} utilizes a discriminator network to distinguish between expert policies with others for imitation learning. IQL \cite{kostrikov2022offline} enables the learned policy to improve substantially over the best behavior in the data through generalization, without ever directly querying a Q-function with unseen actions.
For model-based methods, MOPO \cite{MOPO} learns a pessimistic dynamics model and use it to learn a conservative estimate of the value function; COMBO \cite{COMBO} learns the value function based on both the offline dataset and data generated via model rollouts, and it suppresses the value function on OOD data generated by the model.
Almost all offline RL methods have a similar philosophy: to introduce conservatism or pessimism in the learned policy \cite{offlineRLsurvey}.

There are efforts to conduct offline RL in recommendation \cite{xiao2021general,gao2022cirs,Keeping-recsys,recsys21Pessimistic,10.1145/3289600.3290999}. However, few works explicitly discuss the Matthew effect in recommendation. \citet{10.1145/3289600.3290999} mentioned this effect in their experiment section, but their method is not tailored to overcome this issue.

%% file: sections/03.preliminaries.tex
\section{Empirical Study on Matthew Effect}
\label{sec:mattew_hurt}
We conduct empirical studies in recommendation to show how the Matthew effect affects user satisfaction.
When the Matthew effect is amplified, the recommender will repeatedly recommend the items with dominant categories. To illustrate the long-term effect on user experience, we explore the logs of the KuaiRand-27K video dataset\footnote{\url{https://kuairand.com/}} \cite{gao2022kuairand} and the LFM-1b music dataset\footnote{\url{http://www.cp.jku.at/datasets/LFM-1b/}} \cite{LFM1b}. KuaiRand-27K contains a 23 GB log recording 27,285 users’ 322,278,385 interactions on 32,038,725 videos with 62 categories, which are collected from April 8th, 2022 to May 8th, 2022. LFM-1b contains a 40GB log recording 120,322 users’ 1,088,161,692 listening events on 32,291,134 tracks with 3,190,371 artists, which are fetched from Last.FM in the range from January 2013 to August 2014.

\begin{figure}[!t]
%   \vspace{-2mm}
    \tabcolsep=-2pt
    \centering
    \begin{tabular}{cc}
    \includegraphics[width=0.505\linewidth]{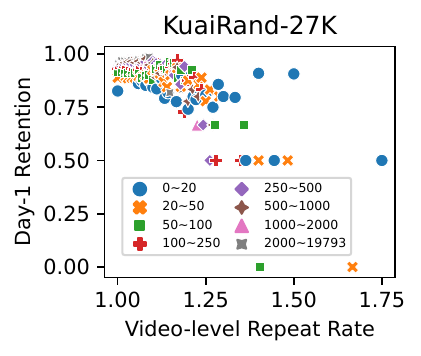} & \includegraphics[width=0.505\linewidth]{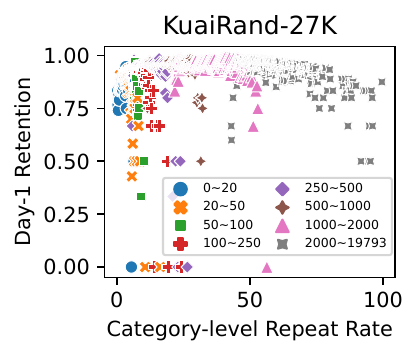}\\
    % \hspace{-3mm}
    \includegraphics[width=0.505\linewidth]{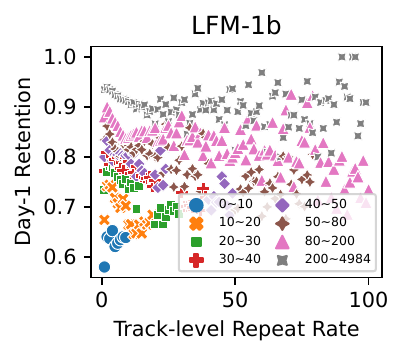} & \includegraphics[width=0.505\linewidth]{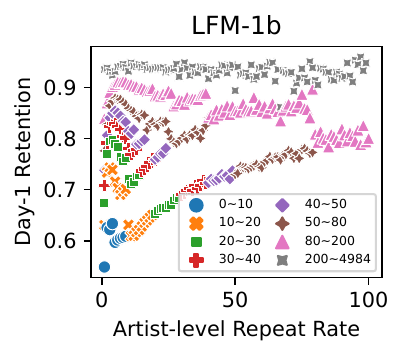}
    % (a) Video-level Overexposure & (b) Category-level Overexposure
    \end{tabular}
    % \vspace{-4mm}
    \caption{
    % The relationship between Day-1 Retention and item-level and category-level repeat rates. 
    An increase in the repeat rate (i.e., the Matthew effect) negatively impacts user experience. In nearly all activity level groups marked by different colors and maker types, Day-1 Retention of users declines as the repeat rate rises.}
    \label{fig:effect}
\end{figure}

Both the two datasets provide the timestamp of each event, hence we can assess the long-term effect of overexposure by investigating the change of Day-1 Retention. Day-1 Retention is defined as the probability of a user who returns to the app tomorrow after finishing today's viewing/listening. This metric is more convincing than real-time signals (e.g., click, adding to favorite) in regard to reflecting the long-term effect on user satisfaction. We consider the item-level and category/artist-level repeat rates as the metrics to measure the Matthew effect. The item-level (or category-level) repeat rate of a user viewing videos on a certain day is defined as:
\begin{equation*}
\frac{\text{the number of viewing events}}{\text{the number of unique videos (or unique categories)}}.
\end{equation*}
For example, if a user views 5 unique videos (which belong to 3 unique categories) 20 times in a day, then the item-level repeat rate is 20/5=4.0 and the category-level repeat rate is 20/3=6.67. The item-level and artist-level repeat rates for music listening are defined in a similar way. Note that in KuaiRand, video-level overexposure rarely appears because of the rule of video recommendation, i.e., the same video will not be recommended twice. 

In this definition, user activity can become a confounder. For instance, a user who views 100 videos a day can be more active than a user who views only 10 videos a day, and thus is more likely to revisit the App the next day. Therefore, we control for this confounder by splitting the users w.r.t. the number of their daily viewing events. Groups with different user activity levels are marked by different colors and marker types. The results are shown in \myfig{effect}. 

In short, Day-1 Retention reduces when the repeat rate increases in each group with a user activity level. This phenomenon can be observed for both the item-level and category/artist-level repeat rates within the video dataset and music dataset. The results show that users' satisfaction will be hurt when the Matthew effect becomes severer.

\section{Preliminary on Model-based RL}
\label{sec:preliminary}
% \subsection{Model-based Reinforcement Learning}

We introduce the basics of RL and model-based offline RL.

% \subsubsection{Basics}
% \smallskip \noindent $\bullet$ \textbf{Basics.}
\subsection{Basics of Reinforcement Learning}
Reinforcement learning (RL) is the science of decision making. We usually formulate the problem as a Markov decision process (MDP): $M = (\mathcal S,\mathcal A,T,r,\gamma)$, where $\mathcal S$ and $\mathcal A$ represent the state space and action space, $T(s,a,s')= P(s_{t+1}=s'|s_t=s,a_t=a)$ is the transition probability from $(s,a)$ to $s'$, $r(s,a)$ is the reward of taking action $a$ at state $s$, and $\gamma$ is the discount factor. Accordingly, the offline MDP can be denoted as $\widehat M=(\mathcal S,\mathcal A,\widehat T,\hat r, \gamma)$, where $\widehat T$ and $\hat r$ are the transition probability and reward function predicted by an offline model. In offline RL, the policy is trained on an offline dataset $\mathcal D$ which was collected by a behavior policy $\pi_\beta$ running in online environment $M$. By modifying the offline MDP $\widehat M$ to be conservative for overcoming the value overestimation issue, we will derive a modified MDP $\widetilde{M}=(\mathcal S,\mathcal A,\widehat T,\widetilde r, \gamma)$, where the modified reward $\widetilde r$ is modified from the predicted reward $\hat r$.

Since RL considers long-term utility, we can define the value function as $V_M^\pi(s) = \mathbb E_{\pi,T}[\sum_{t=0}^{\infty} \gamma^t r(s_t,a_t)|s_0=s]$, denoting the cumulative reward gain by policy $\pi$ after state $s$ in MDP $M$.
Let $P_{T,t}^\pi$ be the probability of the agent's being in state $s$ at time $t$, if the agent uses policy $\pi$ and transits with $T$.
Defining $\rho_{ T}^\pi(s,a) = (1-\gamma)\pi(a|s)\sum_{t=0}^\infty\gamma^t P_{ T,t}^\pi(s)$ as the discounted distribution of state-action pair $(s,a)$ for policy $\pi$ over $T$, we can derive another form of the policy's accumulated reward as $\eta_{ M}(\pi) = \mathbb{E}_{(s,a)\sim\rho_{ T}^\pi}[r(s,a)]$.

% \subsubsection{Model-Based Offline RL Framework}
% \smallskip \noindent $\bullet$ \textbf{Model-Based Offline RL Framework.}
\subsection{Model-Based Offline RL Framework}
In this paper, we follow a state-of-the-art general Model-based Offline Policy Optimization framework, MOPO\footnote{We consider an RL framework to be general if it doesn't require domain-related prior knowledge or any specific algorithms. Satisfying our demands, MOPO is shown to be one of the best-performing model-based offline RL frameworks.} \cite{MOPO}. The basic idea is to learn a dynamics model $\widehat T$ which captures the state transition $(s,a) \rightarrow s'$ of the environment and estimates reward $\hat r(s,a)$ given state $s$ and action $a$.
% we will show how the negative impact of extrapolation error can be mitigated.
For addressing the distributional shift problem where values $V_{\widehat{M}}^\pi(s)$ are usually over-optimistically estimated, MOPO introduces a penalty function $p(s,a)$ on the estimated reward $\hat r(s,a)$ as:
\begin{equation}
\label{eq:mopo_penalty}
    \tilde r(s,a) = \hat r(s,a)-\lambda p(s,a).
\end{equation}
On the modified reward $\tilde r(s,a)$, the offline MDP $\widehat M$ will be modified to be a conservative MDP: $\widetilde{M} = (\mathcal S,\mathcal A,\widehat T,\tilde{r},\gamma)$. 
MOPO learns its policy in this MDP: $\widetilde{M}$. By defining $\epsilon_p(\pi) = \mathbb{E}_{(s,a)\sim\rho_{\widehat T}^\pi}[p(s,a)]$, MOPO has the following theoretical guarantee:
\begin{theorem}
If the penalizer $p(s,a)$ meets:
\begin{equation}
\label{eq:penalty_reg}
    \lambda \mathbb{E}_{(s,a)\sim\rho_{\widehat T}^\pi}[p(s,a)] \ge |\eta_{\widehat{M}}(\pi) - \eta_{M}(\pi)|,
\end{equation}
then the best offline policy $\hat\pi$ trained in $\widetilde M$ satisfies:
\begin{equation}
\label{eq:lower_bound}
    \eta_M(\hat \pi)\ge \sup_\pi\{\eta_M(\pi) - 2\lambda\epsilon_p(\pi)\}.
\end{equation}
\label{theorem:guarantee}
\end{theorem}
% \smallskip
The proof can be found in \cite{MOPO}.
\myeq{penalty_reg} requires the penalty to be a measurement of offline and online mismatch, thus $\epsilon_p(\pi)$ can be interpreted as how much policy $\pi$ will be affected by the offline extrapolation error.
\myeq{lower_bound} is considered to be a theoretical guarantee for reward penalty in model-based offline RL. For example, with $\pi^*$ denoting optimal policy in online MDP $M$, we have $\eta_M(\hat \pi)\ge \eta_M(\pi^*) - 2\lambda\epsilon_p(\pi^*)$.

\smallskip \noindent
\textbf{Remark:} Through learning $\hat\pi$ offline in the conservative MDP $\widetilde{M}$ with \myeq{mopo_penalty}, we can obtain the result that will not deviate too much from the result of learning an optimal policy $\pi^*$ online in the ground-truth MDP $M$. The deviation will not exceed $2\lambda\epsilon_p(\pi^*)$.

However, there was no sufficient analysis on how to properly choose the penalty term $p(s,a)$. Next, We introduce how to adapt this framework to recommendation and reformulate the $p(s,a)$ according to the characteristic of the recommendation scenario.

%% file: sections/04.method.tex
% \section{Preliminaries} 
% \label{sec:method}

\section{Method}
% Here, we incorporate offline model-based RL into recommender system by considering recommendation as a Markov decision process $M$. By using an offline model $\widehat M$ to fit the online MDP $M$, we can train any RL policies in $\widehat M$ rather than online environment $M$, to avoid affecting user experience.

We implement the model-based offline RL framework in recommendation. we redesign the penalty to alleviate the accompanied Matthew effect. Then, we introduce the proposed DORL model.

% While MOPO uses uncertainty as the penalty, we will show this will result in Matthew effect in recommendation.

\subsection{Model-based RL in Recommendation}
In recommendation, we cannot directly obtain a state from the environment, we have to model the state by capturing the interaction context and the user's mood. Usually, a state $s\in \mathcal S$ is defined as the vector extracted from the user's previously interacted items and corresponding feedback. After the system recommends an item as action $a\in \mathcal A$, the user will give feedback as a scale reward signal $r \coloneqq R(s,a)$. For instance, $r\in\{0,1\}$ indicates whether the user clicks the item, or $r\in\mathbb{R}^+$ reflects a user's viewing time for a video. The state transition function (i.e., state encoder) $T$ can be written as $s' \coloneqq f_{\omega}(s,a,r)$, where $f_{\omega}(s,a,r)$ autoregressively outputs the next state $s'$ and can be implemented as any sequential models. 

When learning offline, we cannot obtain users' reactions to the items that are not covered by the offline dataset. we address this problem by using a user model (or reward model) $\widehat R(s,a)$ to learn users' static interests. This model can be implemented as any state-of-the-art recommender such as DeepFM \cite{DeepFM}. The user model will generate an estimated reward $\hat r=\widehat R(s,a)$ representing a user's intrinsic interest in an item. The transition function $\widehat{T}$ will be written as $s' \coloneqq f_{\omega}(s,a,\hat r)$. 
The offline MDP is defined as $\widehat{M} = (\mathcal S,\mathcal A,\widehat T,\hat r,\gamma)$.
% With the settings above, we can extend the previous work \cite{Algorithmic} concerning theoretical analysis on the model's mismatch of model-based RL, and derive our mismatch function as:
Since the estimated reward $\hat r$ can deviate from the ground-truth value $r$, we follow MOPO to use \myeq{mopo_penalty} to get the modified reward $\tilde r(s,a)= \hat r(s,a)-\lambda p(s,a)$. 
Afterward, we can train the recommendation policy on the modified reward $\tilde r$ by treating the user model as simulated users.

Now, the problem turns into designing the penalty term $p(s,a)$. To begin with, we extend the mismatch function in \cite{Algorithmic,MOPO}. We use $R$ and $\widehat R$ as the shorthand for $R(s,a)$ and $\widehat R(s,a)$, respectively.

\begin{lemma}
Define the mismatch function $G_{\widehat M}^\pi(s,a)$ of a policy $\pi$ on the ground truth MDP $M$ and the estimated MDP $\widehat M$ as:
\begin{equation}
\label{eq:G_define}
    \begin{split} \raisetag{3ex}
     &G_{\widehat M}^\pi(s,a) \coloneqq \mathbb E_{\hat s'\sim\widehat{T},\hat r\sim\widehat R}[\gamma V_M^\pi(\hat s') + \hat r] - \mathbb E_{s'\sim T,r\sim R}[\gamma V_M^\pi(s')+r]\\
    & = \mathbb E_{\hat r\sim\widehat R}[\gamma V_M^\pi(f_{\omega}(s,a,\hat r)) + \hat r] - \mathbb E_{r\sim R}[\gamma V_M^\pi(f_{\omega}(s,a,r))+r].
    \end{split}
\end{equation}
It satisfies: %\footnote[2]{The proof of equation(4) is provided in Appendix.}:
\begin{equation}
\label{eq:G_satisfy}
    \mathbb{E}_{(s,a)\sim\rho_{\widehat T}^\pi}[G_{\widehat M}^\pi(s,a)] = \eta_{\widehat{M}}(\pi) - \eta_{M}(\pi).
\end{equation}
\label{lemma:G_p}
\end{lemma}
% \smallskip
% The definition of the mismatch function $G_{\widehat M}^\pi(s,a)$ extends the definition in \cite{MOPO}. Since the state transition is deter
The mismatch function $G_{\widehat M}^\pi(s,a)$ extends the definition presented in \cite{MOPO}, with the key distinction being the separation of state transition function into state $s$ and reward $r$. This is due to the fact that, in the context of recommendation systems, the stochastic nature of state probabilities arises solely from the randomness associated with their reward signals $r$. Consequently, when integrating along the state transition, it is essential to explicitly express the impact of reward $r$.
The proof of \mylemma{G_p} can be adapted from the proof procedure of the telescoping lemma in \cite{Algorithmic}.
 % \myappendix{proof}.

Following the philosophy of conservatism, we add a penalty term $p(s,a)$ according to the mismatch function $G_{\widehat M}^\pi(s,a)$ by assuming:
\begin{equation}
\label{eq:G_penalty}
    \lambda p(s,a) \ge |G_{\widehat M}^\pi(s,a)|.
\end{equation}
By combining \myeq{G_satisfy} and \myeq{G_penalty}, the condition in \myeq{penalty_reg} is met, which provides the theoretical guarantee for the recommendation policy $\pi$ learned in the conservative MDP: $\widetilde{M}$.

\smallskip \noindent
\textbf{Remark:} \mylemma{G_p} provides a perspective for designing the penalty term $p(s,a)$ that satisfies the theoretical guarantee in \mytheorem{guarantee}. According to \myeq{G_penalty}, the problem of defining $p(s,a)$ turns into analyzing $G_{\widehat M}^\pi(s,a)$, which will be described in \mysec{remedy}.

% In order to facilitate the designing of our penalizer function with $|G_{\widehat M}^\pi(s,a)|$, we can split it into two parts:
% \begin{equation}
%     \begin{split}\raisetag{11ex}
%         & |G_{\widehat M}^\pi(s,a)| \\
%        =& |\mathbb E_{\hat r\sim\widehat R(s,a)}[\gamma V_M^\pi(f_{\omega}(s,a,\hat r)) + \hat r] - \mathbb E_{r\sim R(s,a)}[ \gamma V_M^\pi(f_{\omega}(s,a,r))+r]|\\
%         \le& \gamma|\mathbb E_{\hat r\sim\widehat R(s,a)}[V_M^\pi(f_{\omega}(s,a,\hat r))] - \mathbb E_{r\sim R(s,a)}[ V_M^\pi(f_{\omega}(s,a,r))]|\\
%         +& |\mathbb E_{\hat r\sim\widehat R(s,a)}r - \mathbb E_{r\sim R(s,a)}r|\\
%     \text{:= }&\gamma d_V(\widehat{R}(s,a),R(s,a)) + d_1(\widehat{R}(s,a),R(s,a))
%     \end{split}
% \end{equation}

The original MOPO model uses the uncertainty of the dynamics model $P_U$ as the penalty, i.e., $p(s,a)=P_U$. 
% The motivation is that the user model will introduce high variance in the uncertain area, which in turn causes over-optimistic estimation \cite{BEAR}.
However, penalizing uncertainty will encourage the model to pay more attention to items that are frequently recommended while neglecting the rarely recommended ones. This will accelerate the Matthew effect.

\subsection{Matthew Effect}

% \smallskip \noindent $\bullet$ \textbf{Matthew effect in offline RL.}

To quantify the Matthew effect in the results of recommendation, we use a metric: majority category domination (MCD), which is defined as the percentage of the recommended items that are labeled as the dominated categories in training data\footnote{The dominated categories are the most popular categories that cover $80\%$ items in the training set. There are 13 (out of 46) dominated categories in KuaiRand, and 12 (out of 31) dominated categories in KuaiRec.}.

\begin{figure}[!t]
\centering
\includegraphics[width=0.9\linewidth]{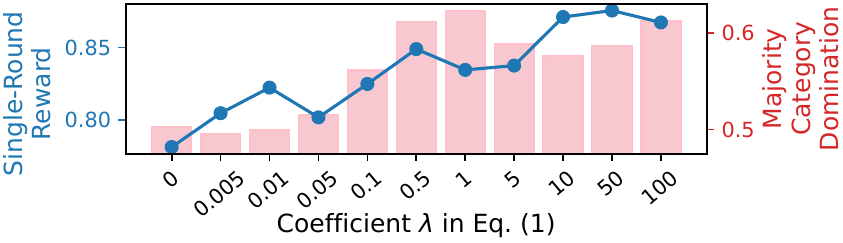}
\vspace{-3mm}
\caption{Effect of conservatism in recommendation.}
% \vspace{-4mm}
% cumulative satisfaction, interaction sequence length, and the reward per turn in
\label{fig:conservative}
\end{figure}

We show the effect of conservatism of MOPO by varying the coefficient $\lambda$ of \myeq{mopo_penalty}. The results on the KuaiRec dataset are shown in \myfig{conservative}. With increasing $\lambda$, the model receives a higher single-round reward (the blue line), which means the policy captures users' interest more accurately. On the other hand, MCD also increases (the red bars), which means the recommended items tend to be the most popular categories (those cover 80\% items) in training data. 
I.e., \textbf{the more conservative the policy is, the stronger the Matthew effect becomes.}

When the results narrow down to these categories, users' satisfaction will be hurt and the interaction process terminates early, which results in low cumulative rewards over the interaction sequence. More details will be described in \mysec{exp}.

\subsection{Solution: Re-design the Penalty}
\label{sec:remedy}
To address this issue, we consider a more sophisticated manner to design the penalty term $p(s,a)$ in \myeq{mopo_penalty}. We dissect the mismatch function in \myeq{G_define} as: 

\begin{equation}
\label{eq:our_G}
    \begin{split} \raisetag{3ex}
        |G_{\widehat M}^\pi(s,a)|  \le & \gamma|\mathbb E_{\hat r\sim\widehat R}[V_M^\pi(f_{\omega}(s,a,\hat r))] - \mathbb E_{r\sim R}[ V_M^\pi(f_{\omega}(s,a,r))]|\\
        & + |\mathbb E_{\hat r\sim\widehat R}r - \mathbb E_{r\sim R}r| \coloneqq \gamma d_V(\widehat{R},R) + d_1(\widehat{R},R),
    \end{split}
\end{equation}
where $d_1(\widehat{R},R)$ represents the deviation of estimated reward $\widehat R$ from true reward $R$, $d_V(\widehat{R},R)$ measures the difference between the value functions $V_M^\pi$ of next state calculated offline (via $\widehat{R}$) and online (via $R$). Both of them can be seen as specific metrics measuring the distance between $\widehat{R}$ and $R$. While $d_1(\widehat{R},R)$ is straightforward, $d_V(\widehat{R},R)$ considers the long-term effect on offline learning and is hard to estimate. 
Based on the aforementioned analysis, a pessimistic reward model $\widehat{R}$ in MOPO will amplify the Matthew effect that reduces long-term satisfaction, thus resulting in a large $d_V(\widehat{R},R)$. 

An intuitive way to solve this dilemma is to introduce exploration on states with high entropy of the logging policy. 
Without access to online user feedback, we can only conduct the counterfactual exploration in the offline data. 

\begin{figure}[!t]
    \centerline{\includegraphics[width=0.98\linewidth]{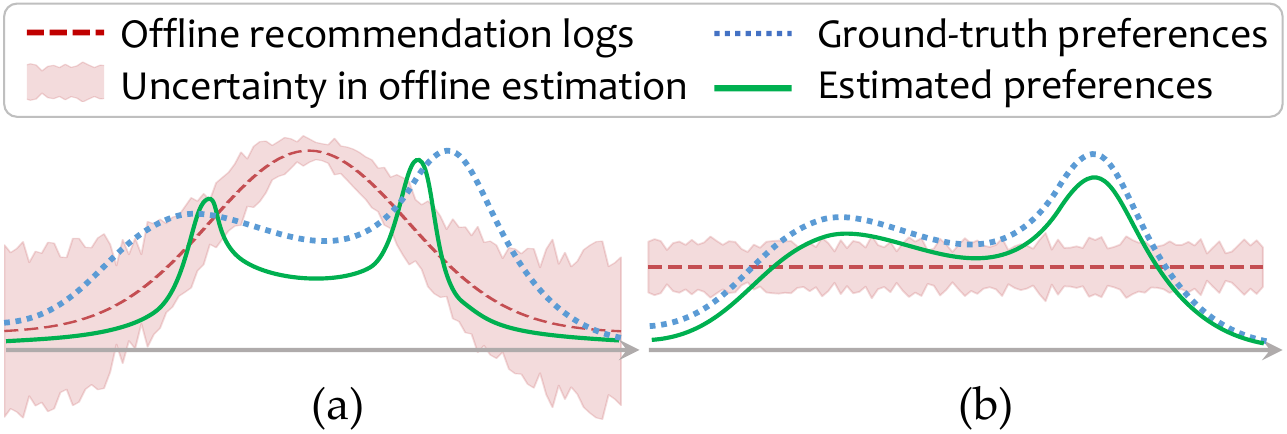}}
    \vspace{-2mm}
    \caption{Estimating user preferences using behavior policies with a Gaussian distribution and a uniform distribution.}
    \label{fig:distribution}
\end{figure}

\smallskip \noindent $\bullet$ \textbf{An illustrative example.}
To illustrate the idea, we give an example in \myfig{distribution}. The goal is to estimate a user's preferences given the logged data induced by a behavior policy. In reality, the distribution of the logged data is dependent on the policies of previous recommenders. For convenience, we use a Gaussian distribution as the behavior policy in \myfig{distribution}(a).
Since previous recommenders cannot precisely reflect users' ground-truth preferences, there is always a deviation between the behavior policy (the red line) and users' ground-truth preferences (the blue line).

Besides, as the items are not equally exposed in the behavior policy, there will be high uncertainty in estimating the rarely appeared items (as shown in the filled area). Offline RL methods emphasize conservatism in estimation and penalize the uncertain samples, which results in the distribution that narrows down the preferences to these dominated items (the green line). This is how the Matthew effect appears.

% \subsection{Matthew Effect Issue}
% \label{sec:matthew_mopo}

% However, estimating the distance $d_V$ given by $V_M^\pi(s,a)$ requires strong assumptions, so we consider an intuitive way here. By analysing the reason why offline model $\widehat{R}$ deviates from $R$, we can find two natural causes:
% \begin{itemize}
    % \item Without sufficient data, predictive model suffers from high variance, which can be measured with model's uncertainty, including epistemic uncertainty and aleatoric uncertainty.
    % \item offline data is collected non-uniformly thus introducing dataset bias on our model, and this is the main factor that brings about \textbf{Matthew effect}: offline recommender only trust those well-covered items, while these top items are selected by a biased behavior policy $\pi_\beta$ and can not reflect the full range of user preferences.
    % \item On one hand, due to insufficient offline data, the user model will introduce high variance in the uncertain area.
% \end{itemize}

% These two aspects can also be observed from \myfig{reason_of_mismatch}:

% \begin{figure}[htbp]
%     \centerline{\includegraphics[width=0.8\linewidth]{figs/show.pdf}}
%     \caption{offline data distribution may deviate from users' real preferences and has high variance for rarely-observed areas.}
%     \label{fig:reason_of_mismatch}
% \end{figure}

By contrast, using a uniform distribution to collect data can prevent biases and reduce uncertainty (\myfig{distribution}(b)). Ideally, a policy learned on sufficient data collected uniformly can capture unbiased user preferences and produce recommendations without the Matthew effect, i.e., $\gamma d_V(\widehat{R},R) + d_1(\widehat{R},R)$ can reduce to $0$. 

Therefore, an intuitive way to design penalty term $p(s,a)$ is to add a term:
the discrepancy of behavior between the uniform distribution $\pi_u(\cdot|s)$ and the behavior policy $\pi_\beta(\cdot|s)$ given state $s$. We use the Kullback–Leibler divergence $D_{KL}(\pi_\beta(\cdot|s)||\pi_u(\cdot|s))$ to measure the distance, which can be written as:
% , as an empirical estimation for  in \myeq{our_G},
% so that our offline policy's performance is improved by alleviating Matthew effect but still keeping conservative about offline data.
% For dataset bias, which is the result of $\pi_\beta$'s non-uniform choice of actions, we can empirically use $D_{KL}(\pi_\beta\|\pi_u)$ as the penalty, where $\pi_u$ chooses action uniformly. With discrete action space size denoted as $|\mathcal{A}|$, we have an entropy penalizer:
\begin{equation}
	\begin{aligned}
		P_E \text{ := } & -D_{KL}(\pi_\beta(\cdot|s)||\pi_u(\cdot|s))\\
		= & -\mathbb E_{a\sim\pi_\beta(\cdot|s)}[\log(\pi_\beta(a|s))-\log(\pi_u(a|s))]\\
		= & ~ \mathcal H(\pi_\beta(\cdot|s)) - \log(|\mathcal A|),
	\end{aligned}
\end{equation}
where $|\mathcal{A}|$ is a constant representing the number of items. Hence, the term $P_E$ depends on the entropy of the behavior policy $\pi_\beta(\cdot|s)$ given state $s$. The modified penalty term can be written as $p(s,a) = P_U + P_E$, and the modified reward model will be formulated as:
\begin{equation}
\label{eq:final_r}
    \tilde{r}(s,a) = \hat r(s,a) - \lambda_1 P_U + \lambda_2 P_E.
\end{equation}
Except for penalizing high uncertainty areas, the new model also penalizes policies with a low entropy at state $s$. Intuitively, if the behavior policy $\pi_\beta(\cdot|s)$ recommended only a few items at state $s$, then the true user preferences at state $s$ may be unrevealed. Under such circumstances, the entropy term $\mathcal H(\pi_\beta(\cdot|s))$ is low, hence we penalize the estimated reward $\hat r(s,a)$ by a large $P_E$.

% Through penalizing entropy, we equivalently encourage large entropy, which helps us achieve counterfactual exploration
The entropy penalizer does not depend on the chosen action but only on the state in which the agent is. Which means the effect of this penalty will be indirect and penalize actions that lead to less diverse states, because of the long-term optimization. Hence, The learned policy achieves the counterfactual exploration in the offline data, which in turn counteracts the Matthew effect in offline RL.

\subsection{The DORL Method}
\begin{figure}[t!]
    \includegraphics[width=0.9\linewidth]{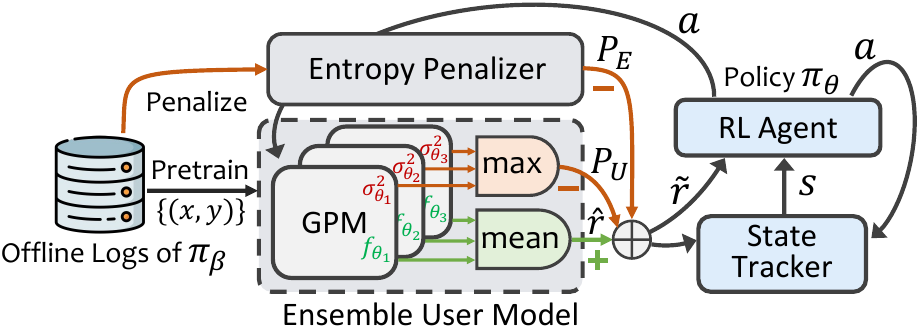}
    \caption{Illustration of Debiased Model-based Offline RL-based Recommendation (DORL) framework.}
    \label{fig:DORL}
\end{figure}

Now, we provide a practical implementation motivated by the analysis above. The proposed model is named Debiased model-based Offline RL (DORL) model, whose framework is illustrated in \myfig{DORL}.

\smallskip \noindent $\bullet$ \textbf{Penalty on entropy.}
We introduce how to compute $P_E$ in the entropy penalizer module. 
\begin{figure}[!t]
    \includegraphics[width=0.9\linewidth]{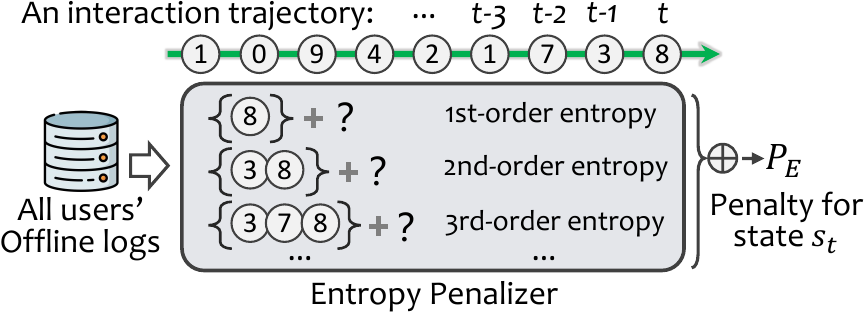}
    \vspace{-2mm}
    \caption{Illustration of the mechanism of entropy penalizer.}
    \label{fig:entropy}
\end{figure}
\myfig{entropy} shows a trajectory of the interaction process, where the current action at time $t$ is to recommend item 8. We define $P_E$ in \myeq{final_r} to be the summation of $k$-order entropy ($k=1,2,\cdots$). For example, when $k=3$, we search all users' recommendation logs to collect all continuous sub-sequences with pattern $[\{3,7,8\},?]$, where ``$?$'' can match any item, and $\{3,7,8\}$ is a sorted set that can cover all of its enumeration, e.g., $[8,3,7]$ or $[7,3,8]$. On these sub-sequences, we can count the frequencies of action ``$?$'' to estimate the entropy of behavior policy $\pi_\beta$ given the previous three recommended items. Without losing generality, we normalize the entropy to range $(0,1]$.

\smallskip \noindent $\bullet$ \textbf{Penalty on uncertainty.}
We penalize both the epistemic uncertainty of the reward model and the aleatoric uncertainty of offline data.
We use the variance of $K$ ensemble reward models $\{\widehat{R}_{\theta_k}, k =1,2,\cdots,K\}$ to capture the epistemic uncertainty, which is commonly used to capture the uncertainty of the model in offline RL \cite{offlineRLsurvey}. Aleatoric uncertainty is data-dependent \cite{TwoUncertainties}. By formulating the user model as a Gaussian probabilistic model (GPM), we can directly predict the variance of the reward and take this predicted variance as aleatoric uncertainty. For the $k$-th model $\widehat{R}_{\theta_k}$, the loss function is:
\begin{equation}
    \mathcal L(\theta_k) = \frac{1}{N}\sum_{i=1}^N\frac{1}{2\sigma^2_{\theta_k}(x_i)}\|y_i-f_{\theta_k}(x_i)\|^2 + \frac{1}{2}\log\sigma^2_{\theta_k}(x_i),
\end{equation}
where $N$ is the number of samples, $f_{\theta_k}(x_i)$ and $\sigma^2_{\theta_k}(x_i)$ are the predicted mean and variance of sample $x_i$, respectively.
By combining epistemic uncertainty and aleatoric uncertainty, we formulate the uncertainty pernalizer $P_U$ in \myeq{final_r} as:
$P_U \text{ := } \max_{k\in \{1,2,\cdots,K\}}\sigma^2_{\theta_k}.$
% \begin{equation}
% P_U \text{ := } \max_{k\in \{1,2,\cdots,K\}}\sigma^2_{\theta_k}(s,a).
% \end{equation}
We define the fitted reward $\hat r$ as the mean of $K$ ensemble models: 
% \begin{equation}
% \hat r(s,a)=\frac{1}{K}\sum_{k}{f_{\theta_k}(s,a)}.
% \end{equation}
$\hat r(s,a)=\frac{1}{K}\sum_{k}{f_{\theta_k}(s,a)}$.
The final modified reward $\tilde{r}(s,a)$ will be computed by \myeq{final_r}. 

The framework of the proposed DORL model is illustrated in \myfig{DORL}. Without losing generality, we use DeepFM \cite{DeepFM} as the backbone for the user model and implement the actor-critic method \cite{a2c} as the RL policy.

The state tracker $f_{\omega}(s,a,r)$ is a network modeling the transition function $T(s,a,s')= P(s_{t+1}=s'|s_t=s,a_t=a)$. It can be implemented as any sequential model such as recurrent neural network (RNN)-based models \cite{wang2022best}, Convolutional models \cite{Caser,NItNet}, Transformer-based methods \cite{sasrec,gao2022cirs,xinxin22}. \citet{jinhuang22} investigated the performances of different state encoders in RL-based recommenders. We use a naive average layer as the state tracker since it requires the least training time but nonetheless outperforms many complex encoders \cite{jinhuang22}. It can be written as:
\begin{equation}
\label{eq:state_tracker}
\vec{s}_{t+1} \coloneqq \frac{1}{N}\sum^{t}_{n=t-N+1}[\vec{e}_{a_n}\oplus \tilde r_n],
\end{equation}
where $\oplus$ is the concatenation symbol, $\vec{s}_{t+1}$ is the vector representing the state at time $t+1$, $\vec{e}_{a_n}$ is the embedding vector of action $a_n$. $\tilde r_n$ is the reward value calculated by \myeq{final_r} and we normalize it to range (0,1] here. $N$ is the window size reflecting how many previous item-reward pairs are calculated.

%% file: sections/05.experiment.tex
\section{Experiments}
\label{sec:exp}

We introduce how we evaluate the proposed DORL model in the interactive recommendation setting. 
We want to investigate the following questions:

\begin{itemize}[leftmargin=*]
    \item \textbf{(RQ1)} How does DORL perform compared to state-of-the-art offline RL methods in the interactive recommendation setting?
    \item \textbf{(RQ2)} To what extent can DORL alleviate the Matthew effect and pursue long-term user experience?
    \item \textbf{(RQ3)} How does DORL perform in different environments with different user tolerance to repeated content?
\end{itemize}

% \noindent
% \textbf{(RQ1)} How does DORL perform compared to state-of-the-art offline RL methods in the interactive recommendation setting?\\
% \textbf{(RQ2)} To what extent can DORL alleviate the Matthew effect and pursue long-term user experience?\\
% \textbf{(RQ3)} How does DORL perform in different environments with different user tolerance to repeated content?\\

% \textbf{(RQ3)} What is the effect of the key parameters in CIRS?

\subsection{Experimental Setup}
We introduce the experimental settings with regard to environments and state-of-the-art offline RL methods.

\subsubsection{Recommendation Environments}
\label{seq:env}

As mentioned in \mysec{IRS}, in the interactive recommendation setting, we are interested in users' long-term satisfaction rather than users' fitting capabilities \cite{deffayet2023offline}.
\textbf{Traditional recommendation datasets are too sparse or lack necessary information }(e.g., timestamps, explicit feedback, item categories) to evaluate the interactive recommender systems.
We create two recommendation environments on two recently-proposed datasets, KuaiRec and KuaiRand-Pure, which contain high-quality logs. 

\noindent\textbf{KuaiRec} \cite{gao2022kuairec} is a video dataset that contains a fully-observed user-item interaction matrix where 1,411 users have viewed all 3,327 videos and left feedback. By taking the fully-observed matrix as users' true interest, we can give a reward for the model's every recommendation (without missing entries like other datasets). We use the normalized viewing time (i.e., the ratio of viewing time to the video length) as the online reward.

\noindent\textbf{KuaiRand-Pure} \cite{gao2022kuairand} is a video dataset that inserted 1,186,059 random recommendations involving 7,583 items into 27,285 users' standard recommendation streams. These randomly exposed data can reflect users' unbiased preferences, from which we can complete the matrix to emulate the fully-observed matrix in KuaiRec. This is an effective way to evaluate RL-based recommendation \cite{jinhuang22,Keeping-recsys}. We use the ``is\_click'' signal to indicate users' ground-truth interest, i.e., as the online reward.

In \mysec{mattew_hurt}, we have shown that users' experience can be hurt by the Matthew effect. To let the environments reflect this phenomenon, we follow \cite{gao2022cirs,10.1145/3511808.3557300} to introduce a quit mechanism: when the model recommends more than $M$ items with the same category in previous $N$ rounds, the interaction terminates. Note that the same item will not be recommended twice in an interaction sequence. Since we evaluate the model via the cumulative rewards $\sum_{t}r_t$ over the interaction trajectory, quitting early (due to the Matthew effect) will lead to inferior performances.
For now, the two environments can play the same role as the online users. Therefore, we can evaluate the model as the process shown in \myfig{settings} (b). 

The evaluation environments are used for assessing models and they are not available in the training stage. For the training purpose, both KuaiRec and KuaiRand provide additional recommendation logs. The statistics of the training data are illustrated in \mytable{data}. 

\begin{table}[t]
\caption{Statistics of two datasets.}
\vspace{-2mm}
\label{tab:data}
\tabcolsep=5pt
\small
\begin{tabular}{@{}cccccc@{}}
\toprule
Datasets                  & Usage & \#Users & \#Items & \#Interactions & \#Categories \\ \midrule
\multirow{2}{*}{KuaiRec}  & Train & 7,176   & 10,728  & 12,530,806     & 31           \\
                          & Test  & 1,411   & 3,327   & 4,676,570      & 31           \\ \cmidrule(lr){1-6}
\multirow{2}{*}{KuaiRand} & Train & 27,285  & 7,551   & 1,436,609      & 46           \\
                          & Test  & 27,285  & 7,583   & 1,186,059      & 46           \\ \bottomrule
\end{tabular}
\vspace{-3mm}
\end{table}

\subsubsection{Baselines}
We select two naive bandit-based algorithms, four model-free offline RL methods, and four model-based offline RL methods (including ours) in evaluation. We use the DeepFM model \cite{DeepFM} as the backbone in the two bandit methods and four model-based methods. These baselines are:
\begin{itemize}[leftmargin=*]
    \item \textbf{$\epsilon$-greedy}, a naive bandit-based policy that outputs a random result with probability $\epsilon$ or outputs the deterministic results of DeepFM with probability $1-\epsilon$.
    \item \textbf{UCB}, a naive bandit-based policy that maintains an upper confidence bound for each item and follows the principle of optimism in the face of uncertainty. %It means if we are uncertain about an action, we should give it a try. UCB can balance the exploration and exploitation in decision-making process.
    \item \textbf{SQN}, or Self-Supervised Q-learning \cite{xinxin20}, contains two output layers (heads): one for the cross-entropy loss and the other for RL. 
    % The RL head serves as a regularizer to introduce users' long-term satisfaction in the self-supervised head. In our interactive setting, 
    We use the RL head to generate final recommendations.
    \item \textbf{BCQ}, or Batch-Constrained deep Q-learning \cite{BCQ}, adapts the conventional deep Q-learning to batch RL. We use the discrete-action version \cite{BCQdiscrete}, whose core idea is to reject these uncertain data and update the policy using only the data of high confidence.
    \item \textbf{CQL}, or Conservative Q-Learning \cite{CQL}, is a model-free RL method that adds a Q-value regularizer on top of an actor-critic policy.% network to avoid overestimating values.
    \item \textbf{CRR}, or Critic Regularized Regression \cite{CRR}, is a model-free RL method that learns the policy by avoiding OOD actions.
    \item \textbf{MBPO}, a vanilla model-based policy optimization method that uses DeepFM as the user model to train an actor-critic policy.% network. % It does not consider conservatism.
    \item \textbf{IPS} \cite{swaminathan2015counterfactual} is a well-known statistical technique adjusting the target distribution by re-weighting each sample in the collected data. 
    % In recommendation, it is widely used for modeling the probability of observation in order to remove the exposure bias or selection bias in the collected data. 
    We implement IPS in a DeepFM-based user model, then learn the policy using an actor-critic method.
    \item \textbf{MOPO}, a model-based offline policy optimization method \cite{MOPO} that penalizes the uncertainty of the DeepFM-based user model and then learns an actor-critic policy.
    % \item \textbf{DORL} is our proposed debiased model-based offline RL model that penalizes both uncertainty and entropy of the user model and learns an actor-critic policy.
\end{itemize}

\subsection{Overall Performance Comparison (RQ1)}
We evaluate all methods in two environments. For the four model-based RL methods (MBPO, IPS, MOPO, and our DORL), we use the same DeepFM model as the user model and fixed its parameters to make sure the difference comes only from the policies. We use the grid search technique on the key parameters to tune all methods in the two environments. For DORL, we search the combination of two key parameters $\lambda_1$ and $\lambda_2$ in \myeq{final_r}. Both of them are searched in $\{0.001, 0.005, 0.01, 0.05, 0.1, 0.5, 1.0, 5, 10, 50, 100\}$. We report the results with $\lambda_1=0.01,\lambda_2=0.05$ for KuaiRand and $\lambda_1=0.05,\lambda_2=5$ for KuaiRec. 
All methods in two environments are evaluated with the quit parameters: $M=0, N=4$, and the maximum round is set to 30. The results are the average metrics of $100$ interaction trajectories.

\begin{figure}[!t]
\centering
\includegraphics[width=1\linewidth]{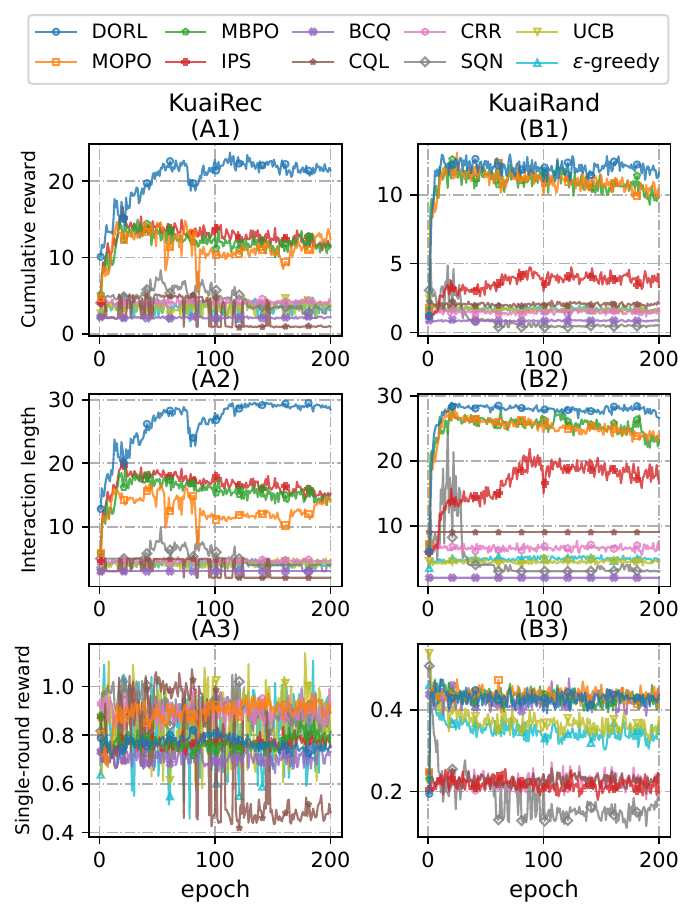}
% \begin{tabular*}{0.8\linewidth}{@{\extracolsep{\fill}}ll}
% (a) KuaiRec & (b) KuaiRand
% \end{tabular*}
\vspace{-3mm}
\caption{Results of all methods in two environments.}
\label{fig:main_result}
\end{figure}

\begin{table*}[t]
\caption{Average Results of all methods in the two environments. (Bold: Best; underline: runner-up). } % MCD is the shorthand for majority category domination. 
\label{tab:results}
% \vspace{-2mm}
\tabcolsep=3pt
% \small
% \footnotesize
\renewcommand\arraystretch{1.1}
\begin{tabular}{lcccccccc}
\toprule
\multirow{2}{*}{Methods} & \multicolumn{4}{c}{KuaiRec} & \multicolumn{4}{c}{KuaiRand} \\ \cmidrule(lr){2-5} \cmidrule(lr){6-9}
 &          $\text{R}_\text{tra}$ &        $\text{R}_\text{each}$ &                         Length &                           MCD &          $\text{R}_\text{tra}$ &        $\text{R}_\text{each}$ &                         Length &                           MCD \\ 
\midrule
UCB               &               $3.606\pm 0.609$ &              $0.853\pm 0.114$ &               $4.219\pm 0.389$ &              $0.811\pm 0.058$ &               $1.651\pm 0.152$ &              $0.372\pm 0.028$ &               $4.431\pm 0.212$ &              $0.789\pm 0.024$ \\
$\epsilon$-greedy &               $3.515\pm 0.731$ &              $0.828\pm 0.129$ &               $4.219\pm 0.405$ &              $0.823\pm 0.048$ &               $1.711\pm 0.126$ &              $0.351\pm 0.025$ &               $4.880\pm 0.270$ &              $0.773\pm 0.024$ \\
SQN               &               $4.673\pm 1.215$ &     $\mathbf{0.913\pm 0.055}$ &               $5.111\pm 1.288$ &              $0.686\pm 0.093$ &               $0.912\pm 0.929$ &              $0.182\pm 0.058$ &               $4.601\pm 3.712$ &              $0.621\pm 0.187$ \\
CRR               &               $4.163\pm 0.253$ &  \underline{$0.895\pm 0.037$} &               $4.654\pm 0.215$ &              $0.865\pm 0.017$ &               $1.481\pm 0.124$ &              $0.226\pm 0.015$ &               $6.561\pm 0.352$ &              $0.733\pm 0.019$ \\
CQL               &               $2.506\pm 1.767$ &              $0.684\pm 0.228$ &               $3.224\pm 1.365$ &              $0.386\pm 0.385$ &               $2.032\pm 0.107$ &              $0.226\pm 0.012$ &               $9.000\pm 0.000$ &              $0.778\pm 0.000$ \\
BCQ               &               $2.123\pm 0.081$ &              $0.708\pm 0.027$ &               $3.000\pm 0.000$ &              $0.667\pm 0.000$ &               $0.852\pm 0.052$ &              $0.425\pm 0.016$ &               $2.005\pm 0.071$ &              $0.998\pm 0.024$ \\
MBPO              &              $12.043\pm 1.312$ &              $0.770\pm 0.029$ &              $15.646\pm 1.637$ &  \underline{$0.362\pm 0.047$} &              $10.933\pm 0.946$ &  \underline{$0.431\pm 0.021$} &  \underline{$25.345\pm 1.819$} &              $0.306\pm 0.040$ \\
IPS               &  \underline{$12.833\pm 1.353$} &              $0.767\pm 0.023$ &  \underline{$16.727\pm 1.683$} &     $\mathbf{0.215\pm 0.064}$ &               $3.629\pm 0.676$ &              $0.216\pm 0.014$ &              $16.821\pm 3.182$ &     $\mathbf{0.201\pm 0.116}$ \\
MOPO              &              $11.427\pm 1.750$ &              $0.892\pm 0.051$ &              $12.809\pm 1.850$ &              $0.479\pm 0.062$ &  \underline{$10.934\pm 0.963$} &     $\mathbf{0.437\pm 0.019}$ &              $25.002\pm 1.891$ &              $0.343\pm 0.029$ \\
Ours              &     $\mathbf{20.494\pm 2.671}$ &              $0.767\pm 0.026$ &     $\mathbf{26.712\pm 3.419}$ &              $0.379\pm 0.015$ &     $\mathbf{11.850\pm 1.036}$ &              $0.428\pm 0.022$ &     $\mathbf{27.609\pm 2.121}$ &  \underline{$0.296\pm 0.036$} \\
\bottomrule
\end{tabular}
\end{table*}

The results are shown in \myfig{main_result}, where all policies are learned with 200 epochs. After learning in each epoch, we will evaluate all methods with $100$ episodes (i.e., interaction trajectories) in the two interactive environments. The first row shows the cumulative reward, which directly reflects the long-term satisfaction in our interactive recommendation setting. The second row and third rows dissect the cumulative reward into two parts: the length of the interaction trajectory and the single-round reward, respectively. For a better comparison, we average the results in 200 epochs and show them in \mytable{results}. Besides the three metrics, we also report the majority category domination (MCD) in \mytable{results}.

From the results, we observe that the four model-based RL methods (MBPO, IPS, MOPO, and DORL) significantly outperform the four model-free RL methods (SQN, CRR, CQL, and BCQ) with respect to trajectory length and cumulative reward. This is because model-based RL is much more sample efficient than model-free RL. In recommendation, the training data is highly sparse. Model-free RL learns directly from the recommendation logs that we have split into different sequences according to the exit rule described above. However, it is extremely difficult to capture the exit mechanism from the sparse logs. By contrast, the model-based RL can leverage the user model to construct as many interaction sequences as possible during training, which guarantees that the policy can distill useful knowledge from the limited offline samples. That is why we embrace model-based RL in recommendation.

\smallskip\noindent\textbf{For model-based RL methods,}
 MOPO shows an obvious improvement compared to the vanilla method MBPO in terms of single-round reward. It is because MBPO does not consider the OOD actions in the offline data that will incur extrapolation errors in the policy. MOPO introduces the uncertainty penalizer to make the policy pay more attention to the samples of high confidence, which in turn makes the policy capture users' interest more precisely. However, MOPO sacrifices many unpopular items because they appear less frequently and are considered uncertain samples. Therefore, the average length decreases, which in turn reduces the cumulative reward. Our method DORL overcomes this problem. From \myfig{main_result}, we observe that DORL attains the maximal average cumulative reward after several epochs in both KuaiRec and KuaiRand due to that it reaches the largest interaction length. Compared to MOPO and MBPO, DORL sacrifices a little bit of the single-round reward due to its counterfactual exploration philosophy, meanwhile, this greatly improves the diversity and enlarges the length of interactions due. Therefore, it achieves the goal of maximizing users' long-term experiences. After enhancing the vanilla MBPO with the IPS technique, the learned user model gives adjustments to the distribution of training data by re-weighting all items. IPS obtains a satisfactory performance in KuaiRec but receives abysmal performances in KuaiRand. This is due to its well-known high variance issue can incur estimation errors. Compared to IPS's hard debiased mechanism, DORL's soft debiased method is more suitable for model-based RL in recommendation.

\smallskip\noindent\textbf{For model-free RL methods,}
As discussed above, four model-free methods fail in two datasets due to limited offline samples. Though they can capture users' interest by returning a high single-round reward (e.g., SQN and CRR in KuaiRec, and BCQ in KuaiRand), they cannot maintain a long interaction trajectory. For example, BCQ updates its policy only on those samples with high confidence, which results in a severe Matthew effect in the recommendation results (reflected by high MCD and short length). SQN's performance oscillates with the largest magnitude since its network is updated by two heads. The RL head serves as a regularizer to the self-supervised head. When the objectives of the two heads conflict with each other, the performance becomes unstable. 
Therefore, these methods are not suitable for recommendation where offline data are sparse.

\smallskip\noindent
As for the naive bandit methods, UCB and $\epsilon$-greedy, they are designed to explore and exploit the optimal actions for the independent and identically distributed (IID) data. They do not even possess the capability to optimize long-term rewards at all. Therefore, they are inclined to recommend the same items when the model finishes exploring offline data, which leads to high MCD and short interactions. These naive policies are not suitable for pursuing long-term user experiences in recommendation.

\subsection{Results on alleviating Matthew effect (RQ2)}

\begin{figure}[!t]
\centering
\includegraphics[width=0.95\linewidth]{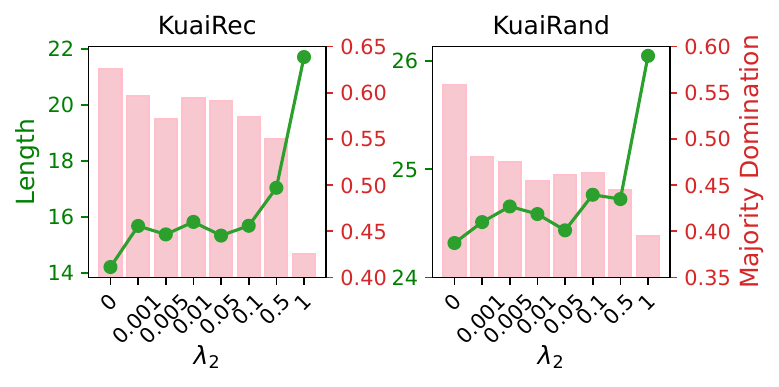}
\vspace{-2mm}
\caption{Effect of penalizing entropy.}
\vspace{-2mm}
% cumulative satisfaction, interaction sequence length, and the reward per turn in
\label{fig:res_entropy}
\end{figure}

We have shown in \myfig{conservative} that penalizing uncertainty will result in the Matthew effect in recommendation. More specifically, increasing $\lambda_1$ can make the recommended items to be the most dominant ones in the training set, which results in a high MCD value.
Here, we show how the introduced ``counterfactual'' exploration mechanism helps alleviate this effect. We conduct the experiments for different combinations of $(\lambda_1, \lambda_2)$ as described above, then we average the results along the $\lambda_1$ to show the influence of $\lambda_2$ alone. 

The results are shown in \myfig{res_entropy}. Obviously, increasing $\lambda_2$ can lengthen the interaction process and reduce majority category domination. I.e., When we penalize the entropy of behavior policy hard,  (1) the recommender does not repeat the items with the same categories; (2) the recommended results will be diverse instead of focusing on dominated items. The results show the effectiveness of penalizing entropy in DORL in alleviating the Matthew effect.

\subsection{Results with different environments (RQ3)}
\begin{figure}[!t]
\centering
\includegraphics[width=0.95\linewidth]{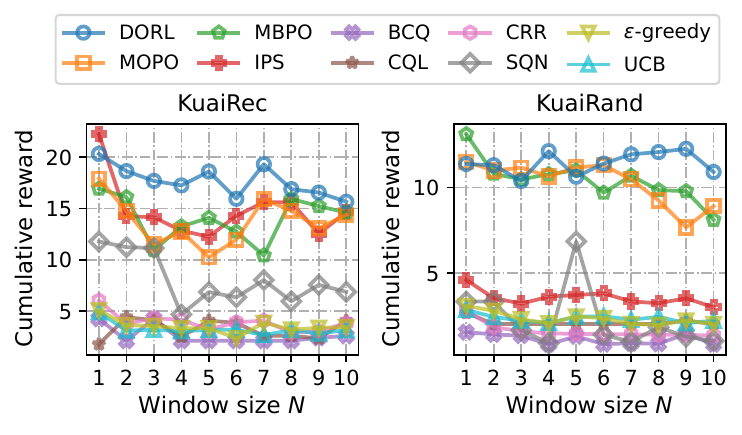}
% \vspace{-4mm}
\caption{Results under different ending conditions}
\vspace{-4mm}
% cumulative satisfaction, interaction sequence length, and the reward per turn in
\label{fig:leave}
\end{figure}

To validate that DORL can work robustly in different environment settings, we vary the window size $N$ in the exit mechanism and fix $M=10$ during the evaluation.  
The results are shown in \myfig{leave}. We only visualize the most important metric: the cumulative reward. When $N$ is small ($N=1$), other model-based methods can surpass our DORL. When $N$ gets larger ($N>3$), users' tolerance for similar content (i.e., items with the same category) becomes lower, and the interaction process comes to be easier to terminate. Under such a circumstance, DORL outperforms all other policies, which demonstrates the robustness of DORL in different environments.

%% file: DORL-final.bbl

\begin{thebibliography}{63}

%%% the Web address.

\ifx \showCODEN    \undefined \def \showCODEN     #1{\unskip}     \fi
\ifx \showDOI      \undefined \def \showDOI       #1{#1}\fi
\ifx \showISBNx    \undefined \def \showISBNx     #1{\unskip}     \fi
\ifx \showISBNxiii \undefined \def \showISBNxiii  #1{\unskip}     \fi
\ifx \showISSN     \undefined \def \showISSN      #1{\unskip}     \fi
\ifx \showLCCN     \undefined \def \showLCCN      #1{\unskip}     \fi
\ifx \shownote     \undefined \def \shownote      #1{#1}          \fi
\ifx \showarticletitle \undefined \def \showarticletitle #1{#1}   \fi
\ifx \showURL      \undefined \def \showURL       {\relax}        \fi
% invisible to TeX
\providecommand\bibfield[2]{#2}
\providecommand\bibinfo[2]{#2}
\providecommand\natexlab[1]{#1}
\providecommand\showeprint[2][]{arXiv:#2}

\bibitem[Afsar et~al\mbox{.}(2022)]%
        {afsar2022reinforcement}
\bibfield{author}{\bibinfo{person}{M~Mehdi Afsar}, \bibinfo{person}{Trafford
  Crump}, {and} \bibinfo{person}{Behrouz Far}.}
  \bibinfo{year}{2022}\natexlab{}.
\newblock \showarticletitle{Reinforcement Learning based Recommender Systems: A
  Survey}.
\newblock \bibinfo{journal}{\emph{Comput. Surveys}} \bibinfo{volume}{55},
  \bibinfo{number}{7} (\bibinfo{year}{2022}), \bibinfo{pages}{1--38}.
\newblock


\bibitem[Agarwal et~al\mbox{.}(2020)]%
        {agarwal2020optimistic}
\bibfield{author}{\bibinfo{person}{Rishabh Agarwal}, \bibinfo{person}{Dale
  Schuurmans}, {and} \bibinfo{person}{Mohammad Norouzi}.}
  \bibinfo{year}{2020}\natexlab{}.
\newblock \showarticletitle{An optimistic perspective on offline reinforcement
  learning}. In \bibinfo{booktitle}{\emph{International Conference on Machine
  Learning}} \emph{(\bibinfo{series}{ICML '20})}. PMLR,
  \bibinfo{pages}{104--114}.
\newblock


\bibitem[Anderson et~al\mbox{.}(2020)]%
        {www20Algorithmic}
\bibfield{author}{\bibinfo{person}{Ashton Anderson}, \bibinfo{person}{Lucas
  Maystre}, \bibinfo{person}{Ian Anderson}, \bibinfo{person}{Rishabh Mehrotra},
  {and} \bibinfo{person}{Mounia Lalmas}.} \bibinfo{year}{2020}\natexlab{}.
\newblock \showarticletitle{Algorithmic Effects on the Diversity of Consumption
  on Spotify}. In \bibinfo{booktitle}{\emph{Proceedings of The Web Conference
  2020}} (Taipei, Taiwan) \emph{(\bibinfo{series}{WWW '20})}.
  \bibinfo{pages}{2155–2165}.
\newblock
\showISBNx{9781450370233}


\bibitem[Cai et~al\mbox{.}(2023a)]%
        {cai2023reinforcing}
\bibfield{author}{\bibinfo{person}{Qingpeng Cai}, \bibinfo{person}{Shuchang
  Liu}, \bibinfo{person}{Xueliang Wang}, \bibinfo{person}{Tianyou Zuo},
  \bibinfo{person}{Wentao Xie}, \bibinfo{person}{Bin Yang},
  \bibinfo{person}{Dong Zheng}, \bibinfo{person}{Peng Jiang}, {and}
  \bibinfo{person}{Kun Gai}.} \bibinfo{year}{2023}\natexlab{a}.
\newblock \showarticletitle{Reinforcing User Retention in a Billion Scale Short
  Video Recommender System}.
\newblock \bibinfo{journal}{\emph{arXiv preprint arXiv:2302.01724}}
  (\bibinfo{year}{2023}).
\newblock


\bibitem[Cai et~al\mbox{.}(2023b)]%
        {cai2023two}
\bibfield{author}{\bibinfo{person}{Qingpeng Cai}, \bibinfo{person}{Zhenghai
  Xue}, \bibinfo{person}{Chi Zhang}, \bibinfo{person}{Wanqi Xue},
  \bibinfo{person}{Shuchang Liu}, \bibinfo{person}{Ruohan Zhan},
  \bibinfo{person}{Xueliang Wang}, \bibinfo{person}{Tianyou Zuo},
  \bibinfo{person}{Wentao Xie}, \bibinfo{person}{Dong Zheng}, {et~al\mbox{.}}}
  \bibinfo{year}{2023}\natexlab{b}.
\newblock \showarticletitle{Two-Stage Constrained Actor-Critic for Short Video
  Recommendation}.
\newblock \bibinfo{journal}{\emph{arXiv preprint arXiv:2302.01680}}
  (\bibinfo{year}{2023}).
\newblock


\bibitem[Chen et~al\mbox{.}(2019)]%
        {10.1145/3289600.3290999}
\bibfield{author}{\bibinfo{person}{Minmin Chen}, \bibinfo{person}{Alex Beutel},
  \bibinfo{person}{Paul Covington}, \bibinfo{person}{Sagar Jain},
  \bibinfo{person}{Francois Belletti}, {and} \bibinfo{person}{Ed~H. Chi}.}
  \bibinfo{year}{2019}\natexlab{}.
\newblock \showarticletitle{Top-K Off-Policy Correction for a REINFORCE
  Recommender System}. In \bibinfo{booktitle}{\emph{Proceedings of the Twelfth
  ACM International Conference on Web Search and Data Mining}} (Melbourne VIC,
  Australia) \emph{(\bibinfo{series}{WSDM '19})}. \bibinfo{pages}{456–464}.
\newblock
\showISBNx{9781450359405}


\bibitem[Chen et~al\mbox{.}(2022)]%
        {minmin22}
\bibfield{author}{\bibinfo{person}{Minmin Chen}, \bibinfo{person}{Can Xu},
  \bibinfo{person}{Vince Gatto}, \bibinfo{person}{Devanshu Jain},
  \bibinfo{person}{Aviral Kumar}, {and} \bibinfo{person}{Ed Chi}.}
  \bibinfo{year}{2022}\natexlab{}.
\newblock \showarticletitle{Off-Policy Actor-Critic for Recommender Systems}.
  In \bibinfo{booktitle}{\emph{Proceedings of the 16th ACM Conference on
  Recommender Systems}} (Seattle, WA, USA) \emph{(\bibinfo{series}{RecSys
  '22})}. \bibinfo{pages}{338–349}.
\newblock
\showISBNx{9781450392785}


\bibitem[Deffayet et~al\mbox{.}(2023)]%
        {deffayet2023offline}
\bibfield{author}{\bibinfo{person}{Romain Deffayet}, \bibinfo{person}{Thibaut
  Thonet}, \bibinfo{person}{Jean-Michel Renders}, {and}
  \bibinfo{person}{Maarten de Rijke}.} \bibinfo{year}{2023}\natexlab{}.
\newblock \showarticletitle{Offline Evaluation for Reinforcement Learning-based
  Recommendation: A Critical Issue and Some Alternatives}.
\newblock \bibinfo{journal}{\emph{arXiv preprint arXiv:2301.00993}}
  (\bibinfo{year}{2023}).
\newblock


\bibitem[Ebert et~al\mbox{.}(2018)]%
        {ebert2018visual}
\bibfield{author}{\bibinfo{person}{Frederik Ebert}, \bibinfo{person}{Chelsea
  Finn}, \bibinfo{person}{Sudeep Dasari}, \bibinfo{person}{Annie Xie},
  \bibinfo{person}{Alex Lee}, {and} \bibinfo{person}{Sergey Levine}.}
  \bibinfo{year}{2018}\natexlab{}.
\newblock \showarticletitle{Visual foresight: Model-based deep reinforcement
  learning for vision-based robotic control}.
\newblock \bibinfo{journal}{\emph{arXiv preprint arXiv:1812.00568}}
  (\bibinfo{year}{2018}).
\newblock


\bibitem[Fujimoto et~al\mbox{.}(2019a)]%
        {BCQdiscrete}
\bibfield{author}{\bibinfo{person}{Scott Fujimoto}, \bibinfo{person}{Edoardo
  Conti}, \bibinfo{person}{Mohammad Ghavamzadeh}, {and} \bibinfo{person}{Joelle
  Pineau}.} \bibinfo{year}{2019}\natexlab{a}.
\newblock \showarticletitle{Benchmarking batch deep reinforcement learning
  algorithms}.
\newblock \bibinfo{journal}{\emph{arXiv preprint arXiv:1910.01708}}
  (\bibinfo{year}{2019}).
\newblock


\bibitem[Fujimoto et~al\mbox{.}(2019b)]%
        {BCQ}
\bibfield{author}{\bibinfo{person}{Scott Fujimoto}, \bibinfo{person}{David
  Meger}, {and} \bibinfo{person}{Doina Precup}.}
  \bibinfo{year}{2019}\natexlab{b}.
\newblock \showarticletitle{Off-Policy Deep Reinforcement Learning without
  Exploration}. In \bibinfo{booktitle}{\emph{Proceedings of the 36th
  International Conference on Machine Learning}}
  \emph{(\bibinfo{series}{Proceedings of Machine Learning Research},
  Vol.~\bibinfo{volume}{97})}, \bibfield{editor}{\bibinfo{person}{Kamalika
  Chaudhuri} {and} \bibinfo{person}{Ruslan Salakhutdinov}} (Eds.).
  \bibinfo{pages}{2052--2062}.
\newblock


\bibitem[Gao et~al\mbox{.}(2022a)]%
        {gao2022cirs}
\bibfield{author}{\bibinfo{person}{Chongming Gao}, \bibinfo{person}{Wenqiang
  Lei}, \bibinfo{person}{Jiawei Chen}, \bibinfo{person}{Shiqi Wang},
  \bibinfo{person}{Xiangnan He}, \bibinfo{person}{Shijun Li},
  \bibinfo{person}{Biao Li}, \bibinfo{person}{Yuan Zhang}, {and}
  \bibinfo{person}{Peng Jiang}.} \bibinfo{year}{2022}\natexlab{a}.
\newblock \showarticletitle{CIRS: Bursting Filter Bubbles by Counterfactual
  Interactive Recommender System}.
\newblock \bibinfo{journal}{\emph{arXiv preprint arXiv:2204.01266}}
  (\bibinfo{year}{2022}).
\newblock


\bibitem[Gao et~al\mbox{.}(2021)]%
        {gao2021advances}
\bibfield{author}{\bibinfo{person}{Chongming Gao}, \bibinfo{person}{Wenqiang
  Lei}, \bibinfo{person}{Xiangnan He}, \bibinfo{person}{Maarten {de Rijke}},
  {and} \bibinfo{person}{Tat-Seng Chua}.} \bibinfo{year}{2021}\natexlab{}.
\newblock \showarticletitle{Advances and Challenges in Conversational
  Recommender Systems: A Survey}.
\newblock \bibinfo{journal}{\emph{AI Open}}  \bibinfo{volume}{2}
  (\bibinfo{year}{2021}), \bibinfo{pages}{100--126}.
\newblock
\showISSN{2666-6510}


\bibitem[Gao et~al\mbox{.}(2022b)]%
        {gao2022kuairec}
\bibfield{author}{\bibinfo{person}{Chongming Gao}, \bibinfo{person}{Shijun Li},
  \bibinfo{person}{Wenqiang Lei}, \bibinfo{person}{Jiawei Chen},
  \bibinfo{person}{Biao Li}, \bibinfo{person}{Peng Jiang},
  \bibinfo{person}{Xiangnan He}, \bibinfo{person}{Jiaxin Mao}, {and}
  \bibinfo{person}{Tat-Seng Chua}.} \bibinfo{year}{2022}\natexlab{b}.
\newblock \showarticletitle{KuaiRec: A Fully-Observed Dataset and Insights for
  Evaluating Recommender Systems}. In \bibinfo{booktitle}{\emph{Proceedings of
  the 31st ACM International Conference on Information \& Knowledge
  Management}} (Atlanta, GA, USA) \emph{(\bibinfo{series}{CIKM '22})}.
  \bibinfo{pages}{540–550}.
\newblock
\urldef\tempurl%
\url{https://doi.org/10.1145/3511808.3557220}
\showDOI{\tempurl}


\bibitem[Gao et~al\mbox{.}(2022c)]%
        {gao2022kuairand}
\bibfield{author}{\bibinfo{person}{Chongming Gao}, \bibinfo{person}{Shijun Li},
  \bibinfo{person}{Yuan Zhang}, \bibinfo{person}{Jiawei Chen},
  \bibinfo{person}{Biao Li}, \bibinfo{person}{Wenqiang Lei},
  \bibinfo{person}{Peng Jiang}, {and} \bibinfo{person}{Xiangnan He}.}
  \bibinfo{year}{2022}\natexlab{c}.
\newblock \showarticletitle{KuaiRand: An Unbiased Sequential Recommendation
  Dataset with Randomly Exposed Videos}. In
  \bibinfo{booktitle}{\emph{Proceedings of the 31st ACM International
  Conference on Information and Knowledge Management}} (Atlanta, GA, USA)
  \emph{(\bibinfo{series}{CIKM '22})}. \bibinfo{pages}{3953–3957}.
\newblock
\urldef\tempurl%
\url{https://doi.org/10.1145/3511808.3557624}
\showDOI{\tempurl}


\bibitem[Gao et~al\mbox{.}(2022d)]%
        {FBsigir22}
\bibfield{author}{\bibinfo{person}{Zhaolin Gao}, \bibinfo{person}{Tianshu
  Shen}, \bibinfo{person}{Zheda Mai}, \bibinfo{person}{Mohamed~Reda
  Bouadjenek}, \bibinfo{person}{Isaac Waller}, \bibinfo{person}{Ashton
  Anderson}, \bibinfo{person}{Ron Bodkin}, {and} \bibinfo{person}{Scott
  Sanner}.} \bibinfo{year}{2022}\natexlab{d}.
\newblock \showarticletitle{Mitigating the Filter Bubble While Maintaining
  Relevance: Targeted Diversification with VAE-Based Recommender Systems}. In
  \bibinfo{booktitle}{\emph{Proceedings of the 45th International ACM SIGIR
  Conference on Research and Development in Information Retrieval}} (Madrid,
  Spain) \emph{(\bibinfo{series}{SIGIR '22})}. \bibinfo{pages}{2524–2531}.
\newblock
\showISBNx{9781450387323}


\bibitem[Ge et~al\mbox{.}(2022)]%
        {Pareto22wsdm}
\bibfield{author}{\bibinfo{person}{Yingqiang Ge}, \bibinfo{person}{Xiaoting
  Zhao}, \bibinfo{person}{Lucia Yu}, \bibinfo{person}{Saurabh Paul},
  \bibinfo{person}{Diane Hu}, \bibinfo{person}{Chu-Cheng Hsieh}, {and}
  \bibinfo{person}{Yongfeng Zhang}.} \bibinfo{year}{2022}\natexlab{}.
\newblock \showarticletitle{Toward Pareto Efficient Fairness-Utility Trade-off
  in Recommendation through Reinforcement Learning}. In
  \bibinfo{booktitle}{\emph{Proceedings of the Fifteenth ACM International
  Conference on Web Search and Data Mining}} (Virtual Event, AZ, USA)
  \emph{(\bibinfo{series}{WSDM '22})}. \bibinfo{pages}{316–324}.
\newblock
\showISBNx{9781450391320}


\bibitem[Gilotte et~al\mbox{.}(2018)]%
        {gilotte2018offline}
\bibfield{author}{\bibinfo{person}{Alexandre Gilotte},
  \bibinfo{person}{Cl\'{e}ment Calauz\`{e}nes}, \bibinfo{person}{Thomas
  Nedelec}, \bibinfo{person}{Alexandre Abraham}, {and} \bibinfo{person}{Simon
  Doll\'{e}}.} \bibinfo{year}{2018}\natexlab{}.
\newblock \showarticletitle{Offline A/B Testing for Recommender Systems}. In
  \bibinfo{booktitle}{\emph{WSDM '18}}. \bibinfo{pages}{198–206}.
\newblock
\showISBNx{9781450355810}


\bibitem[Guo et~al\mbox{.}(2017)]%
        {DeepFM}
\bibfield{author}{\bibinfo{person}{Huifeng Guo}, \bibinfo{person}{Ruiming
  Tang}, \bibinfo{person}{Yunming Ye}, \bibinfo{person}{Zhenguo Li}, {and}
  \bibinfo{person}{Xiuqiang He}.} \bibinfo{year}{2017}\natexlab{}.
\newblock \showarticletitle{DeepFM: A Factorization-Machine Based Neural
  Network for CTR Prediction}. In \bibinfo{booktitle}{\emph{Proceedings of the
  26th International Joint Conference on Artificial Intelligence}} (Melbourne,
  Australia) \emph{(\bibinfo{series}{IJCAI'17})}. \bibinfo{pages}{1725–1731}.
\newblock
\showISBNx{9780999241103}


\bibitem[Hansen et~al\mbox{.}(2021)]%
        {wsdm21diversity}
\bibfield{author}{\bibinfo{person}{Christian Hansen}, \bibinfo{person}{Rishabh
  Mehrotra}, \bibinfo{person}{Casper Hansen}, \bibinfo{person}{Brian Brost},
  \bibinfo{person}{Lucas Maystre}, {and} \bibinfo{person}{Mounia Lalmas}.}
  \bibinfo{year}{2021}\natexlab{}.
\newblock \showarticletitle{Shifting Consumption towards Diverse Content on
  Music Streaming Platforms}. In \bibinfo{booktitle}{\emph{Proceedings of the
  14th ACM International Conference on Web Search and Data Mining}} (Virtual
  Event, Israel) \emph{(\bibinfo{series}{WSDM '21})}.
  \bibinfo{pages}{238–246}.
\newblock
\showISBNx{9781450382977}


\bibitem[Ho and Ermon(2016)]%
        {GAIL}
\bibfield{author}{\bibinfo{person}{Jonathan Ho} {and} \bibinfo{person}{Stefano
  Ermon}.} \bibinfo{year}{2016}\natexlab{}.
\newblock \showarticletitle{Generative Adversarial Imitation Learning}. In
  \bibinfo{booktitle}{\emph{Advances in Neural Information Processing Systems}}
  \emph{(\bibinfo{series}{NeurIPS '16}, Vol.~\bibinfo{volume}{29})},
  \bibfield{editor}{\bibinfo{person}{D.~Lee}, \bibinfo{person}{M.~Sugiyama},
  \bibinfo{person}{U.~Luxburg}, \bibinfo{person}{I.~Guyon}, {and}
  \bibinfo{person}{R.~Garnett}} (Eds.).
\newblock


\bibitem[Huang et~al\mbox{.}(2022)]%
        {jinhuang22}
\bibfield{author}{\bibinfo{person}{Jin Huang}, \bibinfo{person}{Harrie
  Oosterhuis}, \bibinfo{person}{Bunyamin Cetinkaya}, \bibinfo{person}{Thijs
  Rood}, {and} \bibinfo{person}{Maarten de Rijke}.}
  \bibinfo{year}{2022}\natexlab{}.
\newblock \showarticletitle{State Encoders in Reinforcement Learning for
  Recommendation: A Reproducibility Study}. In
  \bibinfo{booktitle}{\emph{Proceedings of the 45th International ACM SIGIR
  Conference on Research and Development in Information Retrieval}} (Madrid,
  Spain) \emph{(\bibinfo{series}{SIGIR '22})}. \bibinfo{pages}{2738–2748}.
\newblock
\showISBNx{9781450387323}


\bibitem[Huang et~al\mbox{.}(2020)]%
        {Keeping-recsys}
\bibfield{author}{\bibinfo{person}{Jin Huang}, \bibinfo{person}{Harrie
  Oosterhuis}, \bibinfo{person}{Maarten de Rijke}, {and} \bibinfo{person}{Herke
  van Hoof}.} \bibinfo{year}{2020}\natexlab{}.
\newblock \showarticletitle{Keeping Dataset Biases out of the Simulation: A
  Debiased Simulator for Reinforcement Learning Based Recommender Systems}. In
  \bibinfo{booktitle}{\emph{RecSys '20}}. \bibinfo{pages}{190–199}.
\newblock
\showISBNx{9781450375832}


\bibitem[Jeunen and Goethals(2021)]%
        {recsys21Pessimistic}
\bibfield{author}{\bibinfo{person}{Olivier Jeunen} {and} \bibinfo{person}{Bart
  Goethals}.} \bibinfo{year}{2021}\natexlab{}.
\newblock \showarticletitle{Pessimistic Reward Models for Off-Policy Learning
  in Recommendation}. In \bibinfo{booktitle}{\emph{Proceedings of the 15th ACM
  Conference on Recommender Systems}} (Amsterdam, Netherlands)
  \emph{(\bibinfo{series}{RecSys '21})}. \bibinfo{pages}{63–74}.
\newblock
\showISBNx{9781450384582}


\bibitem[Kang and McAuley(2018)]%
        {sasrec}
\bibfield{author}{\bibinfo{person}{Wang-Cheng Kang} {and}
  \bibinfo{person}{Julian McAuley}.} \bibinfo{year}{2018}\natexlab{}.
\newblock \showarticletitle{Self-attentive sequential recommendation}. In
  \bibinfo{booktitle}{\emph{International Conference on Data Mining}}
  \emph{(\bibinfo{series}{ICDM '18})}. IEEE, \bibinfo{pages}{197--206}.
\newblock


\bibitem[Kendall and Gal(2017)]%
        {TwoUncertainties}
\bibfield{author}{\bibinfo{person}{Alex Kendall} {and} \bibinfo{person}{Yarin
  Gal}.} \bibinfo{year}{2017}\natexlab{}.
\newblock \showarticletitle{What Uncertainties Do We Need in Bayesian Deep
  Learning for Computer Vision?}. In \bibinfo{booktitle}{\emph{Proceedings of
  the 31st International Conference on Neural Information Processing Systems}}
  (Long Beach, California, USA) \emph{(\bibinfo{series}{NeurIPS '17})}.
  \bibinfo{pages}{5580–5590}.
\newblock
\showISBNx{9781510860964}


\bibitem[Kidambi et~al\mbox{.}(2020)]%
        {MOReL}
\bibfield{author}{\bibinfo{person}{Rahul Kidambi}, \bibinfo{person}{Aravind
  Rajeswaran}, \bibinfo{person}{Praneeth Netrapalli}, {and}
  \bibinfo{person}{Thorsten Joachims}.} \bibinfo{year}{2020}\natexlab{}.
\newblock \showarticletitle{MOReL: Model-Based Offline Reinforcement Learning}.
  In \bibinfo{booktitle}{\emph{Advances in Neural Information Processing
  Systems}} \emph{(\bibinfo{series}{NeurIPS '20}, Vol.~\bibinfo{volume}{33})},
  \bibfield{editor}{\bibinfo{person}{H.~Larochelle},
  \bibinfo{person}{M.~Ranzato}, \bibinfo{person}{R.~Hadsell},
  \bibinfo{person}{M.F. Balcan}, {and} \bibinfo{person}{H.~Lin}} (Eds.).
  \bibinfo{pages}{21810--21823}.
\newblock


\bibitem[Konda and Tsitsiklis(1999)]%
        {a2c}
\bibfield{author}{\bibinfo{person}{Vijay Konda} {and} \bibinfo{person}{John
  Tsitsiklis}.} \bibinfo{year}{1999}\natexlab{}.
\newblock \showarticletitle{Actor-critic algorithms}.
\newblock \bibinfo{journal}{\emph{Advances in neural information processing
  systems}}  \bibinfo{volume}{12} (\bibinfo{year}{1999}).
\newblock


\bibitem[Kostrikov et~al\mbox{.}(2022)]%
        {kostrikov2022offline}
\bibfield{author}{\bibinfo{person}{Ilya Kostrikov}, \bibinfo{person}{Ashvin
  Nair}, {and} \bibinfo{person}{Sergey Levine}.}
  \bibinfo{year}{2022}\natexlab{}.
\newblock \showarticletitle{Offline Reinforcement Learning with Implicit
  Q-Learning}. In \bibinfo{booktitle}{\emph{International Conference on
  Learning Representations}} \emph{(\bibinfo{series}{ICLR '22})}.
\newblock


\bibitem[Kumar et~al\mbox{.}(2019)]%
        {BEAR}
\bibfield{author}{\bibinfo{person}{Aviral Kumar}, \bibinfo{person}{Justin Fu},
  \bibinfo{person}{Matthew Soh}, \bibinfo{person}{George Tucker}, {and}
  \bibinfo{person}{Sergey Levine}.} \bibinfo{year}{2019}\natexlab{}.
\newblock \showarticletitle{Stabilizing off-policy q-learning via bootstrapping
  error reduction}.
\newblock \bibinfo{journal}{\emph{Advances in Neural Information Processing
  Systems}}  \bibinfo{volume}{32} (\bibinfo{year}{2019}).
\newblock


\bibitem[Kumar et~al\mbox{.}(2020)]%
        {CQL}
\bibfield{author}{\bibinfo{person}{Aviral Kumar}, \bibinfo{person}{Aurick
  Zhou}, \bibinfo{person}{George Tucker}, {and} \bibinfo{person}{Sergey
  Levine}.} \bibinfo{year}{2020}\natexlab{}.
\newblock \showarticletitle{Conservative Q-Learning for Offline Reinforcement
  Learning}. In \bibinfo{booktitle}{\emph{Advances in Neural Information
  Processing Systems}} \emph{(\bibinfo{series}{NeurIPS '20},
  Vol.~\bibinfo{volume}{33})},
  \bibfield{editor}{\bibinfo{person}{H.~Larochelle},
  \bibinfo{person}{M.~Ranzato}, \bibinfo{person}{R.~Hadsell},
  \bibinfo{person}{M.F. Balcan}, {and} \bibinfo{person}{H.~Lin}} (Eds.).
  \bibinfo{pages}{1179--1191}.
\newblock


\bibitem[Lei et~al\mbox{.}(2021)]%
        {gao2021tutorial}
\bibfield{author}{\bibinfo{person}{Wenqiang Lei}, \bibinfo{person}{Chongming
  Gao}, {and} \bibinfo{person}{Maarten de Rijke}.}
  \bibinfo{year}{2021}\natexlab{}.
\newblock \showarticletitle{RecSys 2021 Tutorial on Conversational
  Recommendation: Formulation, Methods, and Evaluation}
  \emph{(\bibinfo{series}{RecSys '21})}. \bibinfo{publisher}{Association for
  Computing Machinery}, \bibinfo{address}{New York, NY, USA},
  \bibinfo{pages}{842–844}.
\newblock
\showISBNx{9781450384582}
\urldef\tempurl%
\url{https://doi.org/10.1145/3460231.3473325}
\showDOI{\tempurl}


\bibitem[Levine et~al\mbox{.}(2020)]%
        {offlineRLsurvey}
\bibfield{author}{\bibinfo{person}{Sergey Levine}, \bibinfo{person}{Aviral
  Kumar}, \bibinfo{person}{George Tucker}, {and} \bibinfo{person}{Justin Fu}.}
  \bibinfo{year}{2020}\natexlab{}.
\newblock \showarticletitle{Offline reinforcement learning: Tutorial, review,
  and perspectives on open problems}.
\newblock \bibinfo{journal}{\emph{arXiv preprint arXiv:2005.01643}}
  (\bibinfo{year}{2020}).
\newblock


\bibitem[Liang et~al\mbox{.}(2021)]%
        {diversitySigir21}
\bibfield{author}{\bibinfo{person}{Yile Liang}, \bibinfo{person}{Tieyun Qian},
  \bibinfo{person}{Qing Li}, {and} \bibinfo{person}{Hongzhi Yin}.}
  \bibinfo{year}{2021}\natexlab{}.
\newblock \showarticletitle{Enhancing Domain-Level and User-Level Adaptivity in
  Diversified Recommendation}. In \bibinfo{booktitle}{\emph{Proceedings of the
  44th International ACM SIGIR Conference on Research and Development in
  Information Retrieval}} (Virtual Event, Canada) \emph{(\bibinfo{series}{SIGIR
  '21})}. \bibinfo{pages}{747–756}.
\newblock
\showISBNx{9781450380379}


\bibitem[Liu et~al\mbox{.}(2023)]%
        {liu2023exploration}
\bibfield{author}{\bibinfo{person}{Shuchang Liu}, \bibinfo{person}{Qingpeng
  Cai}, \bibinfo{person}{Bowen Sun}, \bibinfo{person}{Yuhao Wang},
  \bibinfo{person}{Ji Jiang}, \bibinfo{person}{Dong Zheng},
  \bibinfo{person}{Kun Gai}, \bibinfo{person}{Peng Jiang},
  \bibinfo{person}{Xiangyu Zhao}, {and} \bibinfo{person}{Yongfeng Zhang}.}
  \bibinfo{year}{2023}\natexlab{}.
\newblock \showarticletitle{Exploration and Regularization of the Latent Action
  Space in Recommendation}.
\newblock \bibinfo{journal}{\emph{arXiv preprint arXiv:2302.03431}}
  (\bibinfo{year}{2023}).
\newblock


\bibitem[Liu and Huang(2021)]%
        {ExamingMatthew}
\bibfield{author}{\bibinfo{person}{Ying~Chieh Liu} {and}
  \bibinfo{person}{Min~Qi Huang}.} \bibinfo{year}{2021}\natexlab{}.
\newblock \showarticletitle{Examining the Matthew Effect on YouTube
  Recommendation System}. In \bibinfo{booktitle}{\emph{International Conference
  on Technologies and Applications of Artificial Intelligence}}
  \emph{(\bibinfo{series}{TAAI '21})}. \bibinfo{pages}{146--148}.
\newblock
\urldef\tempurl%
\url{https://doi.org/10.1109/TAAI54685.2021.00035}
\showDOI{\tempurl}


\bibitem[Luo et~al\mbox{.}(2018)]%
        {Algorithmic}
\bibfield{author}{\bibinfo{person}{Yuping Luo}, \bibinfo{person}{Huazhe Xu},
  \bibinfo{person}{Yuanzhi Li}, \bibinfo{person}{Yuandong Tian},
  \bibinfo{person}{Trevor Darrell}, {and} \bibinfo{person}{Tengyu Ma}.}
  \bibinfo{year}{2018}\natexlab{}.
\newblock \showarticletitle{Algorithmic framework for model-based deep
  reinforcement learning with theoretical guarantees}.
\newblock \bibinfo{journal}{\emph{arXiv preprint arXiv:1807.03858}}
  (\bibinfo{year}{2018}).
\newblock


\bibitem[Schedl(2016)]%
        {LFM1b}
\bibfield{author}{\bibinfo{person}{Markus Schedl}.}
  \bibinfo{year}{2016}\natexlab{}.
\newblock \showarticletitle{The LFM-1b Dataset for Music Retrieval and
  Recommendation}. In \bibinfo{booktitle}{\emph{Proceedings of the 2016 ACM on
  International Conference on Multimedia Retrieval}} (New York, New York, USA)
  \emph{(\bibinfo{series}{ICMR '16})}. \bibinfo{pages}{103–110}.
\newblock
\showISBNx{9781450343596}


\bibitem[Swaminathan and Joachims(2015)]%
        {swaminathan2015counterfactual}
\bibfield{author}{\bibinfo{person}{Adith Swaminathan} {and}
  \bibinfo{person}{Thorsten Joachims}.} \bibinfo{year}{2015}\natexlab{}.
\newblock \showarticletitle{Counterfactual Risk Minimization: Learning from
  Logged Bandit Feedback}. In \bibinfo{booktitle}{\emph{International
  Conference on Machine Learning}} \emph{(\bibinfo{series}{ICML '15})}. PMLR,
  \bibinfo{pages}{814--823}.
\newblock


\bibitem[Tang and Wang(2018)]%
        {Caser}
\bibfield{author}{\bibinfo{person}{Jiaxi Tang} {and} \bibinfo{person}{Ke
  Wang}.} \bibinfo{year}{2018}\natexlab{}.
\newblock \showarticletitle{Personalized Top-N Sequential Recommendation via
  Convolutional Sequence Embedding}. In \bibinfo{booktitle}{\emph{Proceedings
  of the Eleventh ACM International Conference on Web Search and Data Mining}}
  (Marina Del Rey, CA, USA) \emph{(\bibinfo{series}{WSDM '18})}.
  \bibinfo{pages}{565–573}.
\newblock
\showISBNx{9781450355810}


\bibitem[Tomlein et~al\mbox{.}(2021)]%
        {recsys21best}
\bibfield{author}{\bibinfo{person}{Matus Tomlein}, \bibinfo{person}{Branislav
  Pecher}, \bibinfo{person}{Jakub Simko}, \bibinfo{person}{Ivan Srba},
  \bibinfo{person}{Robert Moro}, \bibinfo{person}{Elena Stefancova},
  \bibinfo{person}{Michal Kompan}, \bibinfo{person}{Andrea Hrckova},
  \bibinfo{person}{Juraj Podrouzek}, {and} \bibinfo{person}{Maria Bielikova}.}
  \bibinfo{year}{2021}\natexlab{}.
\newblock \showarticletitle{An Audit of Misinformation Filter Bubbles on
  YouTube: Bubble Bursting and Recent Behavior Changes}. In
  \bibinfo{booktitle}{\emph{RecSys '21}}. \bibinfo{pages}{1–11}.
\newblock
\showISBNx{9781450384582}


\bibitem[Wang et~al\mbox{.}(2018)]%
        {QuantifyMatthew}
\bibfield{author}{\bibinfo{person}{Hao Wang}, \bibinfo{person}{Zonghu Wang},
  {and} \bibinfo{person}{Weishi Zhang}.} \bibinfo{year}{2018}\natexlab{}.
\newblock \showarticletitle{Quantitative analysis of Matthew effect and
  sparsity problem of recommender systems}. In \bibinfo{booktitle}{\emph{2018
  IEEE 3rd International Conference on Cloud Computing and Big Data Analysis
  (ICCCBDA)}}. \bibinfo{pages}{78--82}.
\newblock
\urldef\tempurl%
\url{https://doi.org/10.1109/ICCCBDA.2018.8386490}
\showDOI{\tempurl}


\bibitem[Wang et~al\mbox{.}(2020a)]%
        {wang2020statistical}
\bibfield{author}{\bibinfo{person}{Ruosong Wang}, \bibinfo{person}{Dean~P
  Foster}, {and} \bibinfo{person}{Sham~M Kakade}.}
  \bibinfo{year}{2020}\natexlab{a}.
\newblock \showarticletitle{What are the statistical limits of offline RL with
  linear function approximation?}
\newblock \bibinfo{journal}{\emph{arXiv preprint arXiv:2010.11895}}
  (\bibinfo{year}{2020}).
\newblock


\bibitem[Wang et~al\mbox{.}(2022a)]%
        {wang2022best}
\bibfield{author}{\bibinfo{person}{Shiqi Wang}, \bibinfo{person}{Chongming
  Gao}, \bibinfo{person}{Min Gao}, \bibinfo{person}{Junliang Yu},
  \bibinfo{person}{Zongwei Wang}, {and} \bibinfo{person}{Hongzhi Yin}.}
  \bibinfo{year}{2022}\natexlab{a}.
\newblock \showarticletitle{Who Are the Best Adopters? User Selection Model for
  Free Trial Item Promotion}.
\newblock \bibinfo{journal}{\emph{IEEE Transactions on Big Data}}
  (\bibinfo{year}{2022}).
\newblock


\bibitem[Wang et~al\mbox{.}(2022b)]%
        {surrogate}
\bibfield{author}{\bibinfo{person}{Yuyan Wang}, \bibinfo{person}{Mohit Sharma},
  \bibinfo{person}{Can Xu}, \bibinfo{person}{Sriraj Badam},
  \bibinfo{person}{Qian Sun}, \bibinfo{person}{Lee Richardson},
  \bibinfo{person}{Lisa Chung}, \bibinfo{person}{Ed~H. Chi}, {and}
  \bibinfo{person}{Minmin Chen}.} \bibinfo{year}{2022}\natexlab{b}.
\newblock \showarticletitle{Surrogate for Long-Term User Experience in
  Recommender Systems}. In \bibinfo{booktitle}{\emph{Proceedings of the 28th
  ACM SIGKDD Conference on Knowledge Discovery and Data Mining}} (Washington
  DC, USA) \emph{(\bibinfo{series}{KDD '22})}. \bibinfo{pages}{4100–4109}.
\newblock
\showISBNx{9781450393850}


\bibitem[Wang et~al\mbox{.}(2020b)]%
        {CRR}
\bibfield{author}{\bibinfo{person}{Ziyu Wang}, \bibinfo{person}{Alexander
  Novikov}, \bibinfo{person}{Konrad Zolna}, \bibinfo{person}{Josh~S Merel},
  \bibinfo{person}{Jost~Tobias Springenberg}, \bibinfo{person}{Scott~E Reed},
  \bibinfo{person}{Bobak Shahriari}, \bibinfo{person}{Noah Siegel},
  \bibinfo{person}{Caglar Gulcehre}, \bibinfo{person}{Nicolas Heess},
  {et~al\mbox{.}}} \bibinfo{year}{2020}\natexlab{b}.
\newblock \showarticletitle{Critic Regularized Regression}.
\newblock \bibinfo{journal}{\emph{Advances in Neural Information Processing
  Systems}}  \bibinfo{volume}{33} (\bibinfo{year}{2020}),
  \bibinfo{pages}{7768--7778}.
\newblock


\bibitem[Watkins and Dayan(1992)]%
        {watkins1992q}
\bibfield{author}{\bibinfo{person}{Christopher~JCH Watkins} {and}
  \bibinfo{person}{Peter Dayan}.} \bibinfo{year}{1992}\natexlab{}.
\newblock \showarticletitle{Q-learning}.
\newblock \bibinfo{journal}{\emph{Machine learning}} \bibinfo{volume}{8},
  \bibinfo{number}{3} (\bibinfo{year}{1992}), \bibinfo{pages}{279--292}.
\newblock


\bibitem[Wu et~al\mbox{.}(2022)]%
        {10.1145/3477495.3531969}
\bibfield{author}{\bibinfo{person}{Junda Wu}, \bibinfo{person}{Zhihui Xie},
  \bibinfo{person}{Tong Yu}, \bibinfo{person}{Handong Zhao},
  \bibinfo{person}{Ruiyi Zhang}, {and} \bibinfo{person}{Shuai Li}.}
  \bibinfo{year}{2022}\natexlab{}.
\newblock \showarticletitle{Dynamics-Aware Adaptation for Reinforcement
  Learning Based Cross-Domain Interactive Recommendation}
  \emph{(\bibinfo{series}{SIGIR '22})}. \bibinfo{pages}{290–300}.
\newblock
\showISBNx{9781450387323}


\bibitem[Xiao and Wang(2021)]%
        {xiao2021general}
\bibfield{author}{\bibinfo{person}{Teng Xiao} {and} \bibinfo{person}{Donglin
  Wang}.} \bibinfo{year}{2021}\natexlab{}.
\newblock \showarticletitle{A general offline reinforcement learning framework
  for interactive recommendation}. In \bibinfo{booktitle}{\emph{Proceedings of
  the AAAI Conference on Artificial Intelligence}} \emph{(\bibinfo{series}{AAAI
  '21}, Vol.~\bibinfo{volume}{35})}. \bibinfo{pages}{4512--4520}.
\newblock


\bibitem[Xin et~al\mbox{.}(2020)]%
        {xinxin20}
\bibfield{author}{\bibinfo{person}{Xin Xin}, \bibinfo{person}{Alexandros
  Karatzoglou}, \bibinfo{person}{Ioannis Arapakis}, {and}
  \bibinfo{person}{Joemon~M. Jose}.} \bibinfo{year}{2020}\natexlab{}.
\newblock \showarticletitle{Self-Supervised Reinforcement Learning for
  Recommender Systems}. In \bibinfo{booktitle}{\emph{Proceedings of the 43rd
  International ACM SIGIR Conference on Research and Development in Information
  Retrieval}} \emph{(\bibinfo{series}{SIGIR '20})}. \bibinfo{pages}{931–940}.
\newblock
\showISBNx{9781450380164}


\bibitem[Xin et~al\mbox{.}(2022)]%
        {xinxin22}
\bibfield{author}{\bibinfo{person}{Xin Xin}, \bibinfo{person}{Tiago Pimentel},
  \bibinfo{person}{Alexandros Karatzoglou}, \bibinfo{person}{Pengjie Ren},
  \bibinfo{person}{Konstantina Christakopoulou}, {and}
  \bibinfo{person}{Zhaochun Ren}.} \bibinfo{year}{2022}\natexlab{}.
\newblock \showarticletitle{Rethinking Reinforcement Learning for
  Recommendation: A Prompt Perspective}. In
  \bibinfo{booktitle}{\emph{Proceedings of the 45th International ACM SIGIR
  Conference on Research and Development in Information Retrieval}} (Madrid,
  Spain) \emph{(\bibinfo{series}{SIGIR '22})}. \bibinfo{pages}{1347–1357}.
\newblock
\showISBNx{9781450387323}


\bibitem[Xu et~al\mbox{.}(2022)]%
        {10.1145/3511808.3557300}
\bibfield{author}{\bibinfo{person}{Shuyuan Xu}, \bibinfo{person}{Juntao Tan},
  \bibinfo{person}{Zuohui Fu}, \bibinfo{person}{Jianchao Ji},
  \bibinfo{person}{Shelby Heinecke}, {and} \bibinfo{person}{Yongfeng Zhang}.}
  \bibinfo{year}{2022}\natexlab{}.
\newblock \showarticletitle{Dynamic Causal Collaborative Filtering}. In
  \bibinfo{booktitle}{\emph{Proceedings of the 31st ACM International
  Conference on Information \& Knowledge Management}} (Atlanta, GA, USA)
  \emph{(\bibinfo{series}{CIKM '22})}. \bibinfo{pages}{2301–2310}.
\newblock
\showISBNx{9781450392365}


\bibitem[Xue et~al\mbox{.}(2023)]%
        {ResAct}
\bibfield{author}{\bibinfo{person}{Wanqi Xue}, \bibinfo{person}{Qingpeng Cai},
  \bibinfo{person}{Ruohan Zhan}, \bibinfo{person}{Dong Zheng},
  \bibinfo{person}{Peng Jiang}, {and} \bibinfo{person}{Bo An}.}
  \bibinfo{year}{2023}\natexlab{}.
\newblock \showarticletitle{ResAct: Reinforcing Long-term Engagement in
  Sequential Recommendation with Residual Actor} \emph{(\bibinfo{series}{ICLR
  '23})}.
\newblock


\bibitem[Yu et~al\mbox{.}(2021)]%
        {COMBO}
\bibfield{author}{\bibinfo{person}{Tianhe Yu}, \bibinfo{person}{Aviral Kumar},
  \bibinfo{person}{Rafael Rafailov}, \bibinfo{person}{Aravind Rajeswaran},
  \bibinfo{person}{Sergey Levine}, {and} \bibinfo{person}{Chelsea Finn}.}
  \bibinfo{year}{2021}\natexlab{}.
\newblock \showarticletitle{Combo: Conservative offline model-based policy
  optimization}.
\newblock \bibinfo{journal}{\emph{Advances in Neural Information Processing
  Systems}}  \bibinfo{volume}{34} (\bibinfo{year}{2021}),
  \bibinfo{pages}{28954--28967}.
\newblock


\bibitem[Yu et~al\mbox{.}(2020)]%
        {MOPO}
\bibfield{author}{\bibinfo{person}{Tianhe Yu}, \bibinfo{person}{Garrett
  Thomas}, \bibinfo{person}{Lantao Yu}, \bibinfo{person}{Stefano Ermon},
  \bibinfo{person}{James~Y Zou}, \bibinfo{person}{Sergey Levine},
  \bibinfo{person}{Chelsea Finn}, {and} \bibinfo{person}{Tengyu Ma}.}
  \bibinfo{year}{2020}\natexlab{}.
\newblock \showarticletitle{MOPO: Model-based Offline Policy Optimization}. In
  \bibinfo{booktitle}{\emph{Advances in Neural Information Processing Systems}}
  \emph{(\bibinfo{series}{NeurIPS '20}, Vol.~\bibinfo{volume}{33})},
  \bibfield{editor}{\bibinfo{person}{H.~Larochelle},
  \bibinfo{person}{M.~Ranzato}, \bibinfo{person}{R.~Hadsell},
  \bibinfo{person}{M.F. Balcan}, {and} \bibinfo{person}{H.~Lin}} (Eds.).
  \bibinfo{pages}{14129--14142}.
\newblock


\bibitem[Yuan et~al\mbox{.}(2019)]%
        {NItNet}
\bibfield{author}{\bibinfo{person}{Fajie Yuan}, \bibinfo{person}{Alexandros
  Karatzoglou}, \bibinfo{person}{Ioannis Arapakis}, \bibinfo{person}{Joemon~M.
  Jose}, {and} \bibinfo{person}{Xiangnan He}.} \bibinfo{year}{2019}\natexlab{}.
\newblock \showarticletitle{A Simple Convolutional Generative Network for Next
  Item Recommendation}. In \bibinfo{booktitle}{\emph{Proceedings of the Twelfth
  ACM International Conference on Web Search and Data Mining}} (Melbourne VIC,
  Australia) \emph{(\bibinfo{series}{WSDM '19})}. \bibinfo{pages}{582–590}.
\newblock
\showISBNx{9781450359405}


\bibitem[Zhang et~al\mbox{.}(2022)]%
        {10.1145/3534678.3539040}
\bibfield{author}{\bibinfo{person}{Qihua Zhang}, \bibinfo{person}{Junning Liu},
  \bibinfo{person}{Yuzhuo Dai}, \bibinfo{person}{Yiyan Qi},
  \bibinfo{person}{Yifan Yuan}, \bibinfo{person}{Kunlun Zheng},
  \bibinfo{person}{Fan Huang}, {and} \bibinfo{person}{Xianfeng Tan}.}
  \bibinfo{year}{2022}\natexlab{}.
\newblock \showarticletitle{Multi-Task Fusion via Reinforcement Learning for
  Long-Term User Satisfaction in Recommender Systems}. In
  \bibinfo{booktitle}{\emph{Proceedings of the 28th ACM SIGKDD Conference on
  Knowledge Discovery and Data Mining}} (Washington DC, USA)
  \emph{(\bibinfo{series}{KDD '22})}. \bibinfo{pages}{4510–4520}.
\newblock
\showISBNx{9781450393850}


\bibitem[Zhang et~al\mbox{.}(2023)]%
        {zhang2023divide}
\bibfield{author}{\bibinfo{person}{Yuan Zhang}, \bibinfo{person}{Xue Dong},
  \bibinfo{person}{Weijie Ding}, \bibinfo{person}{Biao Li},
  \bibinfo{person}{Peng Jiang}, {and} \bibinfo{person}{Kun Gai}.}
  \bibinfo{year}{2023}\natexlab{}.
\newblock \showarticletitle{Divide and Conquer: Towards Better Embedding-based
  Retrieval for Recommender Systems From a Multi-task Perspective}.
\newblock \bibinfo{journal}{\emph{arXiv preprint arXiv:2302.02657}}
  (\bibinfo{year}{2023}).
\newblock


\bibitem[Zhao et~al\mbox{.}(2021)]%
        {xiangyuwww21}
\bibfield{author}{\bibinfo{person}{Xiangyu Zhao}, \bibinfo{person}{Long Xia},
  \bibinfo{person}{Lixin Zou}, \bibinfo{person}{Hui Liu},
  \bibinfo{person}{Dawei Yin}, {and} \bibinfo{person}{Jiliang Tang}.}
  \bibinfo{year}{2021}\natexlab{}.
\newblock \showarticletitle{UserSim: User Simulation via Supervised Generative
  Adversarial Network}. In \bibinfo{booktitle}{\emph{WWW '21}}.
  \bibinfo{pages}{3582–3589}.
\newblock
\showISBNx{9781450383127}


\bibitem[Zheng et~al\mbox{.}(2021a)]%
        {DGCN}
\bibfield{author}{\bibinfo{person}{Yu Zheng}, \bibinfo{person}{Chen Gao},
  \bibinfo{person}{Liang Chen}, \bibinfo{person}{Depeng Jin}, {and}
  \bibinfo{person}{Yong Li}.} \bibinfo{year}{2021}\natexlab{a}.
\newblock \showarticletitle{DGCN: Diversified Recommendation with Graph
  Convolutional Networks}. In \bibinfo{booktitle}{\emph{Proceedings of the Web
  Conference 2021}} (Ljubljana, Slovenia) \emph{(\bibinfo{series}{WWW '21})}.
  \bibinfo{pages}{401–412}.
\newblock
\showISBNx{9781450383127}


\bibitem[Zheng et~al\mbox{.}(2021b)]%
        {DICE}
\bibfield{author}{\bibinfo{person}{Yu Zheng}, \bibinfo{person}{Chen Gao},
  \bibinfo{person}{Xiang Li}, \bibinfo{person}{Xiangnan He},
  \bibinfo{person}{Yong Li}, {and} \bibinfo{person}{Depeng Jin}.}
  \bibinfo{year}{2021}\natexlab{b}.
\newblock \showarticletitle{Disentangling User Interest and Conformity for
  Recommendation with Causal Embedding}. In
  \bibinfo{booktitle}{\emph{Proceedings of the Web Conference 2021}}
  (Ljubljana, Slovenia) \emph{(\bibinfo{series}{WWW '21})}.
  \bibinfo{pages}{2980–2991}.
\newblock
\showISBNx{9781450383127}


\bibitem[Zou et~al\mbox{.}(2019)]%
        {lixin19}
\bibfield{author}{\bibinfo{person}{Lixin Zou}, \bibinfo{person}{Long Xia},
  \bibinfo{person}{Zhuoye Ding}, \bibinfo{person}{Jiaxing Song},
  \bibinfo{person}{Weidong Liu}, {and} \bibinfo{person}{Dawei Yin}.}
  \bibinfo{year}{2019}\natexlab{}.
\newblock \showarticletitle{Reinforcement Learning to Optimize Long-Term User
  Engagement in Recommender Systems}. In \bibinfo{booktitle}{\emph{Proceedings
  of the 25th ACM SIGKDD International Conference on Knowledge Discovery \&
  Data Mining}} (Anchorage, AK, USA) \emph{(\bibinfo{series}{KDD '19})}.
  \bibinfo{pages}{2810–2818}.
\newblock
\showISBNx{9781450362016}


\bibitem[Zou et~al\mbox{.}(2020)]%
        {Pseudo-Dyna-Q}
\bibfield{author}{\bibinfo{person}{Lixin Zou}, \bibinfo{person}{Long Xia},
  \bibinfo{person}{Pan Du}, \bibinfo{person}{Zhuo Zhang}, \bibinfo{person}{Ting
  Bai}, \bibinfo{person}{Weidong Liu}, \bibinfo{person}{Jian-Yun Nie}, {and}
  \bibinfo{person}{Dawei Yin}.} \bibinfo{year}{2020}\natexlab{}.
\newblock \showarticletitle{Pseudo Dyna-Q: A Reinforcement Learning Framework
  for Interactive Recommendation}. In \bibinfo{booktitle}{\emph{WSDM '20}}.
  \bibinfo{pages}{816–824}.
\newblock
\showISBNx{9781450368223}


\end{thebibliography}
